Review article

# Latent variable models in the era of industrial big data: Extension and beyond

Xiangyin Kong, Xiaoyu Jiang, Bingxin Zhang, Jinsong Yuan, Zhiqiang Ge *

*State Key Laboratory of Industrial Control Technology, College of Control Science and Engineering, Zhejiang University, Hangzhou 310027, Zhejiang, PR China*

## ARTICLE INFO

## ABSTRACT

A rich supply of data and innovative algorithms have made data-driven modeling a popular technique in modern industry. Among various data-driven methods, latent variable models (LVMs) and their counterparts account for a major share and play a vital role in many industrial modeling areas. LVM can be generally divided into statistical learning-based classic LVM and neural networks-based deep LVM (DLVM). We first discuss the definitions, theories and applications of classic LVMs in detail, which serves as both a comprehensive tutorial and a brief application survey on classic LVMs. Then we present a thorough introduction to current mainstream DLVMs with emphasis on their theories and model architectures, soon afterwards provide a detailed survey on industrial applications of DLVMs. The aforementioned two types of LVM have obvious advantages and disadvantages. Specifically, classic LVMs have concise principles and good interpretability, but their model capacity cannot address complicated tasks. Neural networks-based DLVMs have sufficient model capacity to achieve satisfactory performance in complex scenarios, but it comes at sacrifices in model interpretability and efficiency. Aiming at combining the virtues and mitigating the drawbacks of these two types of LVMs, as well as exploring non-neural-network manners to build deep models, we propose a novel concept called lightweight deep LVM (LDLVM). After proposing this new idea, the article first elaborates the motivation and connotation of LDLVM, then provides two novel LDLVMs, along with thorough descriptions on their principles, architectures and merits. Finally, outlooks and opportunities are discussed, including important open questions and possible research directions.

## 1. Introduction

Since the large-scale use of steam engines started in the 1780s opened the prelude to the Industrial Revolution, industry and manufacturing have played a progressively important role in people's daily life (Landes, 2003). Nowadays, modern industry has entered the era of Industry 4.0 (Schwab, 2017; Zhong, Xu, Klotz, & Newman, 2017). Intelligent algorithms and models in industry have great significance in reducing production cost, improving production quality, ensuring plant safety, and promoting smart manufacturing (Ge, Song, Ding, & Huang, 2017; Kusiak, 2019; Tao, Qi, Liu, & Kusiak, 2018). Previously, understanding industrial process usually need to rely on prior knowledge or experiences from experts or engineers. Due to the advances in measurement and information technologies and the wide use of data storage systems, modern industry has entered the era of big data, which means a large amount of data have been recorded and collected in the industrial process (Kong & Ge, 2022a; Shang & You, 2019; Zhu, Ge, Song, & Gao, 2018). A more detailed introduction to the industrial big data is provided in Section 2.2. The plentiful supply of data and the emerging novel algorithms make data-driven modeling a popular tool for understanding industrial process in the past two decades (Geng & Xie, 2019; Jiang & Ge, 2022; Jiang, Yin, Dong, & Kaynak, 2021; Qin, 2012; Tidriri, Chatti, Verron, & Tiplica, 2016; Yang & Ge, 2022; Yao, Shen, Cui, Zheng, & Ge, 2022; Zhuo, Yin, & Ge, 2022). Data-driven modeling methods first explicitly or implicitly extract useful information from industrial data, which is then transferred to effective knowledge for process understanding and decision making (Ge, 2018).

In the past two decades, a lot of data-driven modeling and analyzing methods have been proposed for different industrial applications, such as semiconductor manufacturing (Fan, Hsu, Tsai, He, & Cheng, 2020; Lee, Yang, Lee, & Kim, 2019), injection molding (Guo et al., 2019; Kumar, Park, & Lee, 2020), smart grid (Mohan, Soman, & Kumar, 2018; Xu et al., 2020), wind energy forecasting (Liu & Chen, 2019; Liu, Chen, Lv, Wu, & Liu, 2019), predictive maintenance (Zhang, Yang, & Wang, 2019; Zonta et al., 2020), prognostics health management (Lee, Jin, Liu, & Ardakani, 2017; Tsui, Chen, Zhou, Hai, & Wang, 2015; Watanabe, Sebem, Leal, & Hounsell, 2021) and process industry (Qin, 2012; Severson, Chaiwatanodom, & Braatz, 2016; Sun & Ge, 2021; Zeng & Ge, 2021). Among various data-driven methods, *latent variable*

* Corresponding author.

*E-mail addresses:* xiangyinkong@zju.edu.cn (X. Kong), jiangxiaoyu@zju.edu.cn (X. Jiang), zhangbingxin@zju.edu.cn (B. Zhang), jinsongyuan@zju.edu.cn (J. Yuan), gezhiqiang@zju.edu.cn (Z. Ge).

*models (LVMs)* and their variants account for a large proportion and play a significant role in many application scenarios (Blei, 2014; Ge, 2018; MacGregor, Bruwer, Miletic, Cardin, & Liu, 2015; Qin, Dong, Zhu, Wang, & Liu, 2020; Shao, Han, Li, Ge, & Zhang, 2022; Tomba, Facco, Bezzo, & Barolo, 2013; Yang & Ge, 2020). A detailed description of the significance of LVM for industrial data is presented in Section 2.3, and a comprehensive introduction to the definition of LVM is given in Section 3.1. Generally speaking, LVM is a kind of data analytics method that can *find the underlying infrastructure in data and extract useful information from data*. Such characteristic of LVM makes it especially useful in *dimension reduction*, which perfectly meets the high-dimensional nature of industrial data, where the data sources usually come from hundreds of sensors, and there is usually no obvious spatial or sequential relationship between these sensors. By projecting the data into a low-dimensional latent space, continuous LVMs such as principle component analysis (PCA), independent component analysis (ICA) and partial least squares (PLS) are able to learn pivotal features from the original industrial data and at the same time enhance the accuracy and efficiency of the data analysis process. By decomposing the complex unknown data distribution into multiple simple and known distributions, discrete LVMs such as the Gaussian mixture model (GMM) can effectively model the intricate industrial process in an efficient and clear way. The principles of most LVMs are clear and concise, which means they usually have *good interpretability, few parameters, simple training processes* and *short training time*. For the past two decades, LVMs have been widely used in industrial data clustering, monitoring, diagnosis, visualization, regression, classification (Jing & Hou, 2015; Joswiak, Peng, Castillo, & Chiang, 2019; Messaoud, Bradai, & Moulay, 2019), etc.

However, with the development of industrial process, the traditional LVMs is gradually not competent for the modeling tasks. The main changes include *the scale of production expands significantly, the industrial process becomes more complex and changes more frequently, the accumulation of industrial data is growing rapidly and continuously*, etc. Unfortunately, the model complexity of traditional LVMs is not enough to meet these changes so that they cannot model the current industrial scenarios quite well. Coincidentally, the machine learning (ML) community is also confronted with similar problems during the same period. Before 2006, the ML community is dominated by *statistical learning* models such as *support vector machine* and *decision trees*. But after 2010, neural networks-based *deep learning* (DL) has penetrated most of the researches in ML community.

There are two main *objective* drivers that have promoted the vigorous development of DL: (1) the breakthrough in computing power has laid a solid foundation for the successful training of deep neural networks (DNN); (2) the large amounts of data accumulated in modern society can effectively train DNN to avoid serious overfitting. On the other hand, the rapid development of DL is also inseparable from the community's *subjective* efforts. In order to better train deep networks, researchers have developed many effective training tricks, such as *pre-training and fine-tuning* (Hinton, Osindero, & Teh, 2006; Hinton & Salakhutdinov, 2006), *residual connection* (He, Zhang, Ren, & Sun, 2016), *dropout* (Srivastava, Hinton, Krizhevsky, Sutskever, & Salakhutdinov, 2014), and so on. Due to these positive effects, the "emerging" DL has quickly made remarkable achievements in many fields. Such as convolutional neural networks (CNN) in image recognition (Rawat & Wang, 2017), recurrent neural networks (RNN), long short-term memory networks (LSTM) (Hochreiter & Schmidhuber, 1997) and Transformer (Vaswani, Shazeer, Parmar, Uszkoreit, Jones, Gomez, Kaiser, & Polosukhin, 2017) in natural language processing (Devlin, Chang, Lee, & Toutanova, 2019; Sutskever, Martens, & Hinton, 2011; Sutskever, Vinyals, & Le, 2014), deep generative models like generative adversarial network (GAN) in data generation (Goodfellow et al., 2014), and deep reinforcement learning (DRL) in games (Silver et al., 2016). DL harnesses the hierarchical and cascaded deep model structure to increase model complexity to improve the ability of representation and feature extraction, which is crucial for addressing large datasets and complicated tasks. On the other hand, the nonlinear activation function and sufficient number of neurons and layers enable DNN to approximate arbitrarily complex functions (Nielsen, 2015). DNN also has unique advantages in dealing with unstructured data due to its capability of extracting spatial or temporal information.

The above features of DL make it especially suitable for modeling modern industrial scenarios, where the volume and dimension of data are both quite large, meanwhile the process usually presents strong nonlinearity and dynamics. The industry community has already developed many DL-based methods for modeling (Al-Garadi et al., 2020; Cheng, Lin, Wu, Zhu, & Shao, 2021; Deng, Shao, Hu, Jiang, & Jiang, 2020; He, Zhang, & Zhang, 2022; Jiang & Ge, 2021a; Li et al., 2022; Liu, Miao, Jiang, & Chen, 2020; Shen, Yao, & Ge, 2022; Sun & Ge, 2021; Yang, Li, et al., 2020). Among them, there is a type of widely used deep model that can be considered as an extension of the traditional LVMs, for which we named it as *deep latent variable models (DLVMs)*. Typical DLVMs include deep belief network (DBN), autoencoders (AE) and its variants, e.g., denoising AE (DAE), stacked AE (SAE), variational AE (VAE), and so on. Detailed descriptions and comparisons about current DLVMs are provided in Section 4.1. Benefit from the powerful representation ability of DL, the performance of DLVMs has surpassed a lot of traditional LVMs in various industrial modeling tasks, such as fault diagnosis, prognostics health management and soft sensor.

Existing DLVMs are all neural networks-based deep models, they can be summarized as a kind of DNN model. Anyone who has trained DNN models can feel that it requires a lot of effort to adjust parameters, and the entire training process is quite complex and time-consuming, and sometimes difficult to converge. Actually, the existing DNN-based DLVM does improve the modeling performance, but it also brings other problems that cannot be ignored, such as *poor interpretability, a requirement for large data amounts, hard to design and tune parameters, heavy computation complexity and difficult to train*, etc. Each of these problems is currently one of the hottest frontiers in the DL community. So far, for data-driven industrial process modeling, we have found two interesting and contradictory phenomena: (1) Traditional LVMs have *better model interpretability, fewer parameters, shorter running time*, and they are not severely dependent on the data volume, but their performance may not cope with complex industrial process modeling. (2) Existing neural networks-based DLVMs can achieve better results on many complex industrial modeling tasks, but it also comes at sacrifices in model *interpretability* and *efficiency*, meanwhile they rely too much on the data volume, which means their performance may plummet when the sample size is relatively small.

Is there a way to neutralize the characteristics of the aforesaid two kinds of LVMs? Or in other words, is there a way to build a *high-performance* LVM meanwhile maintaining *good interpretability* and *high modeling efficiency*? As mentioned, DLVMs use the hierarchical model structure to increase model complexity so that to improve the ability of representation and feature extraction, which is crucial for improving performance. But high model complexity also brings a lot of defects, such as poor interpretability and long training time. Looking back at the development of the ML community, it seems that model complexity is crucial to performance improvement, but is contradictory to model interpretability and efficiency. Inspired by the merits and demerits of traditional LVMs and existing DLVMs, in the trade-off between model interpretability and model performance, we provide our own insights in this paper. The development of ML has shown that model complexity (or capacity) is necessary for model performance, and the unique hierarchical and cascaded model structure of DL is a good way to increase complexity. If the hierarchical model structure of current DL models can be applied to traditional LVMs, then can such an application not only improve the performance of traditional LVMs, but also simultaneously maintain the virtues of LVMs themselves (such as good interpretability, convenient training procedures, etc.)? Based on this idea, we propose a new kind of LVM called lightweight deep latent variable model (LDLVM),[1] whose definition and connotation are further analyzed in Section 5.1. Moreover, we provide two novel LDLVMs and detailedly describe their principles, structures and advantages in Sections 5.2 and 5.3, respectively. Generally, the motivations of LDLVM can be summarized as *(1) combining the virtues of traditional LVMs and DLVMs* and *(2) exploring non-neural-network manners to build deep models*. The purpose of LDLVM is to achieve competitive model performance compared with existing DL models in appropriate application scenarios, while overcome some shortcomings of current deep models simultaneously.

This paper focuses on the theme: *latent variable models and their industrial applications*. We discuss three types of LVM, including classic LVM, deep LVM and lightweight deep LVM, as well as their industrial applications in detail. Among them, lightweight deep LVM is a novel latent variable model aiming at overcoming some of the limitations of traditional LVM, deep LVM, and current neural networks-based deep models.

The rest of this article is organized as follows. Section 2 starts with an overview on the background of modern industry, machine learning and deep learning. Section 3 provides a detailed explanation on the definition, theory, and application of classic LVMs. Section 4 begins with a thorough introduction to current mainstream DLVMs, then presents a detailed survey on their industrial applications. Section 5 first discusses the motivation and connotation of LDLVM, then provides two novel LDLVMs, along with thorough descriptions of their principles, structures and merits. Outlooks and opportunities in this field are discussed in Section 6. Finally, Section 7 concludes this article.

## 2. Background

This section briefly introduces the background of modern industry and industrial big data, then discusses the significance of LVMs for industrial data, and finally presents an overview of ML and DL.

### 2.1. Modern industry

After centuries of development, industrial production has evolved a lot. Compared with the past, modern industry has some obvious characteristics:

1. *The scale of industrial production expands significantly:* The production scale of industry has expanded greatly in the past century. In the United States, the *Board of Governors of the Federal Reserve System* has established an indicator to measure the scale of industrial production called the *Industrial Production Index (INDPRO)*. INDPRO is an economic indicator that measures real output for all facilities located in the United States manufacturing, mining, and electric, and gas utilities (Board of Governors of the Federal Reserve System (US), 2021). If we record the value of INDPRO in 1919 as 100, then the change of INDPRO from 1919 to 2019 can be found in Fig. 1(a) (Board of Governors of the Federal Reserve System (US), 2021). The INDPRO of 2019 has increased by 1964% compared to that of 1919.

2. *The industrial process becomes more complex and changes more frequently:* The huge scale of production makes the industrial process more complex and changing more frequently than decades ago. At the same time, the increase in customer demands and the disturbances from external environment may further enhance the complexity of the industrial process. For instance, a change in feedstock or product grade or even changes such as the diurnal load variation of a power plant or the summer–winter operation of a refinery, may cause the industrial process to run under multiple modes (Souza & Araújo, 2014). Moreover, the variation in production process and operating mode is an inherent feature in most industrial applications. For example, in the chemical process, the equipment characteristics will change due to catalyst deactivation, scale adhesion, preventive maintenance and other reasons (Sun & Ge, 2021). The variation of load or seasonal adjustment will also have influence on the whole production. On the other hand, the complex process mechanism also increases the difficulty of modeling modern industry, e.g., in the penicillin fermentation process, microorganisms will experience multiple growth phases (Jin, Chen, Wang, Yang, & Wu, 2016).

3. *The accumulation of industrial data is growing rapidly and continuously:* With the growth of industrial production scale and process complexity, the accumulation of industrial data has also reached a new level. Due to the advances in measurement technologies, various sensors are deployed in the industry to measure real-time status of different variables, such as pressure, temperature, flow, etc. After decades of development, the number of sensors in industry has increased a lot and is still growing. According to China's industry consulting services company, the size and forecast of the global sensor market from 2010 to 2024 are shown in Fig. 1(b). Besides, the large-scale use of storage systems, such as edge devices, distributed control systems and cloud servers, provides a prerequisite for utilizing massive industrial data. The measurement and storage of enormous data have brought modern industry into the era of big data (Qin, 2014). In the meantime, the data form also evolves a lot — from univariate to multivariate, to high-dimensional; from homogeneous data to heterogeneous datasets; from structured to unstructured; from static to dynamic (Sun & Ge, 2021), etc. It is worth noting that, although modern industry has accumulated massive data, in some cases, there may not be enough samples that can be used for modeling. For example, in the field of fault diagnosis, important information is usually contained in fault samples, but some types of fault samples are difficult to collect (Jiang & Ge, 2020; Zhuo & Ge, 2021). Besides, the data that most affects the modeling performance is the *labeled samples*, which may also be hard to obtain under certain circumstances (Moreira de Lima & Ugulino de Araújo, 2021; Yao & Ge, 2017a).

### 2.2. Industrial big data

With the development of modern industry, its accumulated data is also evolving. There are three main types of industrial big data, namely *production data*, *management data* and *external data*. Production data comes from the materials, equipment, and products in the industrial process, which is collected by instruments and sensors. Management data comes from enterprise informatization, including enterprise resource planning (ERP), product life management (PLM), supply chain management (SCM), etc. This type of data is the traditional information asset of industrial enterprises. External data refers to information from outside the enterprise, such as economy, market, customers, government, etc. Among these three types of data, since the production data directly represents the working status of the industrial process, it contains the most information and has the greatest meaning. In a narrow sense, industrial big data can be regarded as production data, which is also the most studied data type in the industrial community. This article also focuses on this kind of data. Unless otherwise specified, the industrial big data in this article refers to production data.

However, it is not easy to fully understand and make good use of industrial big data, especially the production data. In addition to the general features of big data, industrial data also has some unique characteristics, which are discussed as follows:

---

[1] LDLVM is quite similar to the other concept "lightweight deep model (LDM)" that we proposed in Ref. Kong and Ge (2022c). Strictly speaking, LDLVM can be understood as a way to realize LDM, or a subclass of LDM. At present, for narrative convenience, LDLVM in this paper and LDM in Ref. Kong and Ge (2022c) can be considered synonymous.

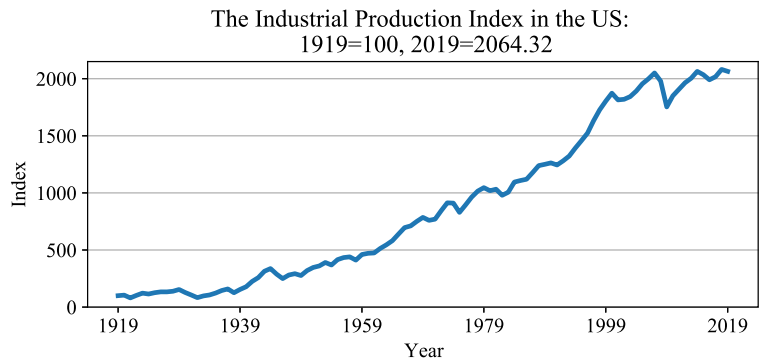

(a) The change of Industrial Production Index (INDPRO) from 1919 to 2019 in the US. The INDPRO value in 1919 is used as baseline, which is set to 100.

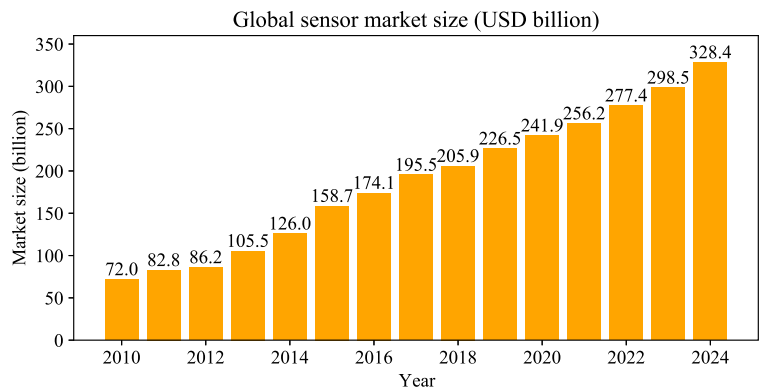

(b) 2010-2024 global sensor market size and forecast.

**Fig. 1.** Some characteristics of modern industry.

- **High dimension:** The increase of the number of instruments and sensors have brought industrial process many new variables, which inevitably leads to the high-dimension nature of industrial big data.
- **Big volume:** The advances in information technology allow sensors to collect data in seconds or even milliseconds level. The rapid measurement and storage of data results in a substantial increase in the volume of data. It is estimated that the data volume of a large industrial enterprise is petabyte or even exabyte level.
- **Nonlinearity:** Since industrial production involves numerous physical transformations and chemical reactions, the processes are usually nonlinear.
- **Dynamicity:** Measurement data is collected continuously in the industrial process, which means industrial data is usually time-series. The data of the current moment is related to the data of the previous moments.
- **Time latency:** Another temporal characteristic of industrial data is long time latency. Input materials need to go through multiple procedures and transformations to become outputs. Therefore, there is a lag between input variables, intermediate variables, and output variables.
- **Multi-mode:** Due to the change and adjustment of the load, the working conditions of industrial production often contain multiple modes. The wide variation between different modalities brings additional difficulties to the analyses of industrial data.
- **Uncertainty:** Environmental noise, equipment noise and measurement noise inevitably exist in the industrial process, which causes almost all industrial data to have varying degrees of uncertainty. In addition, the measurement circumstance in industry is quite harsh. Missing values also widely exist in the measurement variables, which aggravates the uncertainty of industrial data.
- **Semi-supervised:** With the increase of industrial sensors, easily measured variables such as temperature and flow are collected in large quantities. However, due to the expensive measurement costs, some hard-to-measure variables, such as reactant concentrations, are still quite difficult to obtain. These variables are usually the most important variables in the production process and are therefore often used as prediction labels for ML and DL models. This means that the data volume of training data is large, but the sample size of training labels is small, which leads to the natural semi-supervised characteristic of industrial data.

### 2.3. The significance of LVMs for industrial big data

Compared to other models, LVMs have many properties that are quite suitable for industrial scenarios, which make them popular tools for modeling and analyzing industrial data.

- **Dimension reduction and information extraction:** Most continuous LVMs are natural dimension reduction methods. They project the raw data into a low-dimensional latent space, extracting useful information and pivotal features while reducing the data dimension. Since the data sources of modern industry usually come from hundreds of correlative sensors, dimension reduction is quite helpful for improving the accuracy and efficiency of industrial data modeling.
- **Generative hypothesis:** Almost all LVMs have generative assumptions about the original data. They assume that the observed raw variables are redundant and are actually generated by a series of latent variables (this will be detailed in Section 3.1). This perfectly fits the characteristics of industrial variables, where the measurement sensors are often correlated and redundant. For example, engineers may place several temperature sensors in close proximity to monitor an important temperature indicator more accurately. Researchers generally believe that high-dimensional industrial processes are driven by a few core variables, and the generative hypotheses of LVMs meet this requirement.

· **Probabilistic interpretation:** Compared with other models, LVMs have good probabilistic interpretation. The probabilistic definition of LVM decomposes the complex raw data distribution into a series of conditional distributions over the latent variables (see Section 3.1 for more details). The probabilistic perspective of LVMs makes it quite advantageous to deal with data uncertainty, especially the measurement noises and the missing values, which is common in industrial scenarios. Additionally, probabilistic LVMs and can also be easily extended to mixture model form.

· **Graph interpretation:** LVM also has good graph interpretation (Bishop, 2007; Ge, 2018), which is closely related to its generation hypothesis and probability interpretation. The graph interpretation of LVM causes it intimately associated with causal discovery (Bongers, Forré, Peters, & Mooij, 2021; Louizos et al., 2017), which is quite useful for modeling the relationship between industrial process variables. Besides, graphical explanation also makes LVM more intuitive and understandable.

· **Convenient training process:** The training process of most traditional LVMs is quite simple, straightforward and does not take too much time, which is very suitable to serve as preliminary or baseline models for industrial process modeling.

### 2.4. Overview of machine learning and deep learning

Generally, LVMs are a kind of ML algorithm, while DLVMs can be regarded as a type of DL model. To better understand LVMs, the concepts of ML and DL are first introduced briefly.

ML can generally be divided into four paradigms: supervised learning, unsupervised learning, semi-supervised learning, and reinforcement learning.[2] Supervised learning uses the labeled data to learn a function that maps the input to output. Unsupervised learning algorithms self-discover natural patterns from unlabeled data. Most traditional LVMs, such as PCA, ICA and GMM, are unsupervised algorithms. Semi-supervised learning falls between unsupervised learning and supervised learning, which is an approach that combines a small amount of labeled data with a large amount of unlabeled data to train model. Reinforcement learning studies how intelligent agents ought to take actions in an environment in order to maximize the notion of cumulative reward. The features of the above four branches are shown in Fig. 2.

DL can be considered as a powerful kind of ML model, and it evolves from the artificial neural network (ANN). The fundamental element in a neural network (NN) is the artificial neuron. In DNN, there are multiple layers, with each layer consisting of multiple neurons. DL has made remarkable achievements in many fields, which has been outlined in the introduction. At present, when "DL" is used as a term, it is almost synonymous with "DNN". The abovementioned typical DLVMs such as DBN and VAE are actually a class of DNN or DL-based methods. Despite the significant achievements of DNN in many fields, it also brings some new problems, specifically as follows:

1. Interpretability problem: DNN is just like a "black box" and has no gratifying interpretability. We do not really know the whole operation process inside of DNN and how it gives correct results, which usually means there are inevitable concerns about reliability in the practical application of DL.

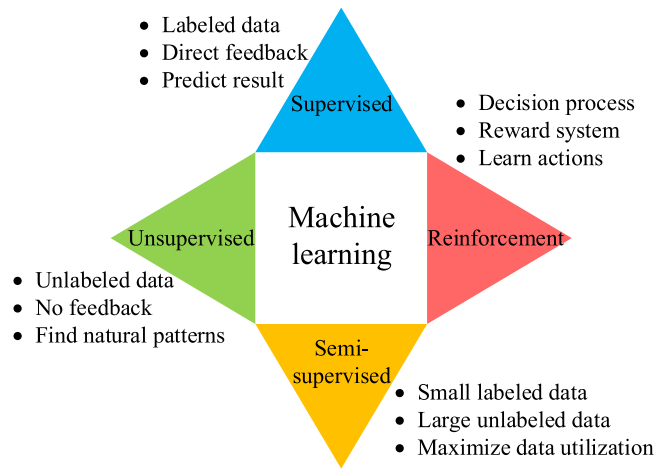

**Fig. 2.** The characteristics of different ML branches.

2. Requirement for large data amounts: DNN needs a lot of training data samples. When the training samples are small (this problem is indeed existing in many practical fields), this modeling method usually leads to intolerably poor performance.
3. Hard to design and tune parameters: DNN usually consists of many model hyperparameters, including many network structure hyperparameters and optimizer hyperparameters. These parameters are quite tricky to assign and adjust.
4. Heavy computation complexity: Massive network weights and biases of DNN need to be updated iteratively during the training phase through backpropagation, which leads to a quite complicated training process, resulting in heavy computation complexity and long training time.

Aiming at solving or alleviating the aforementioned problems to a certain extent, the LDLVM introduced in Section 5.1 can be regarded as a new kind of DL method, which explores non-neural-network manners to build deep models. The fundamental modeling tools of LDLVM are not artificial neurons trained by backpropagation. Instead, LDLVM makes use of simpler approaches, such like the concise LVMs, to construct deep models. More details about LDLVM will be elaborated on Section 5.

## 3. Traditional LVMs

This section discusses the definition, theory and application of classic LVMs in detail, which can be seen as both a tutorial and a brief application survey on traditional LVMs.

### 3.1. Definition: What is an LVM?

To begin with, we first give an introduction to the definition of LVM. A *latent* or *hidden* variable is a variable which is not directly observable. In contrast, a variable that can be directly observed is called *observed* or *manifest* variable. Historically, the idea of latent variables arose primarily from psychometrics (Carreira-Perpinan, 2001; Spearman, 1904), and it now can be understood from different angles. Strictly speaking, an LVM is defined from the perspective of probability. When the observed variables are too complex to model or analyze, we can define a joint distribution over manifest and latent variables, and the corresponding distribution of the observed variables is then obtained by marginalization. This allows relatively complex distributions to be expressed in terms of more tractable joint distributions over the expanded variable space (Bishop, 1998).

Specifically, suppose there is a set of observed variables $x = \{x_1, x_2, \ldots, x_m\}$ whose distribution $p(x)$ is difficult to model. The goal of an LVM is to express the distribution $p(x)$ in terms of a smaller number of latent variables $t = \{t_1, t_2, \ldots, t_k\}$ where $k \leq m$. This can be achieved by first defining a conditional distribution $p(x|t)$ of the data variables given the latent variables, so that the joint distribution is

$$p(x, t) = p(x|t)p(t) \tag{1}$$

where the prior distribution $p(t)$ is usually obey to some assumptions. The desired model for the distribution $p(x)$ can be obtained by marginalizing over the latent variables

$$p(x) = \int p(x|t)p(t)dt. \tag{2}$$

---

[2] In some literature, ML is only divided into two types: supervised learning and unsupervised learning, or into three types: supervised learning, unsupervised learning, and reinforcement learning.

Eqs. (1) and (2) provide a mainstream definition of latent variables in the statistics community. It is clear that integration (2) cannot be directly solved unless some assumptions about $p(t)$ and $p(x|t)$ are provided. A crucial assumption in LVM is the *conditional independence* (Everett, 2013), which can be expressed as

$$p(x|t) = p(x_1|t)p(x_2|t)\ldots p(x_m|t) = \prod_{j=1}^{m} p(x_j|t). \tag{3}$$

Other necessary assumptions include that the distributions $p(t)$ and $p(x|t)$ should be explicitly or implicitly of known form, and dependent on a set of unknown parameters (Everett, 2013). For example, one of the simplest LVMs — factor analysis (FA) assumes the distribution $p(t)$ to be a Gaussian distribution. Although the probabilistic definition of LVM is quite simple and clear, the strict assumptions in this definition may be too difficult to meet in some situations, which may limit the practical applications of LVM.

On the other hand, in the long historical development of LVM, there is also a non-probabilistic, intuitive and mild way to understand this concept. As mentioned, latent variables arose primarily from psychology and sociology. In these areas, latent variables have a more straight connotation, which can be expressed as

**Definition 1.** Latent variables are factors that cannot be directly measured but are assumed to be the underlying generators of observed variables.[3]

Although latent variables are not observable, certain of their effects on manifest variables are observable. For example, direct measurement of a concept such as *racial prejudice* is not possible, however, one could observe whether a person approves or disapproves of a particular piece of government legislation to make a judgement about his/ her level of *racial prejudice* (Everett, 2013). A more detailed and vivid illustration of the connotation of latent variable is presented in Fig. 3 through an example of *China's high school education system*.

The most well-known LVM for now may be the principal component analysis (PCA), however, in the early stage of PCA development, it is traditionally not considered as an LVM. There are two commonly used definitions of PCA, which can be expressed as *the orthogonal projection that maximizes the variance of the projected data* (Hotelling, 1933), or equivalently, *the linear projection that minimizes the average projection cost* (Pearson, 1901). These two definitions finally yield to a same eigenvalue decomposition problem of the data covariance matrix (Bishop, 2007). Traditionally, PCA is essentially a matrix factorization algorithm, and its decomposition does not involve probabilistic modeling at all. So, according to the statistical definition of LVM, PCA cannot be regarded as an LVM. But on the other hand, the principle of PCA assumes that the *projected features* are the fundamental factors that generate the observed data, and this presumption coincides with Definition 1. Moreover, Tipping and Bishop (Tipping & Bishop, 1999) proposed a probabilistic form of the original PCA model in 1999, thus providing a probabilistic explanation for this factorization algorithm and making PCA a formal LVM statistically. PCA has shown effectiveness and unique advantages in a wealth of areas such as chemometrics, economics, biology, medical science, social statistics, and due to the wide application of PCA, it has become a representative method of LVM in the past two decades (Aït-Sahalia & Xiu, 2019; Baker, Bloom, & Davis, 2016; Kolenikov & Angeles, 2009; Martis, Acharya, Mandana, Ray, & Chakraborty, 2012; Song et al., 2006).

There are several other statistical learning algorithms that share similar connotations and principles to PCA. Some of them have been closely related to the concept of latent variables since they were proposed, such as partial least squares (PLS) (also known as projection to latent structures), some of them were only used in a few special fields in the early days of their invention, such as independent component analysis (ICA) for blind source separation. But for now, these methods are all considered as typical traditional LVMs, and we will present a more thorough introduction to them in Section 3.2. Besides, the above methods are all continuous LVMs, in which the extracted latent factors are all continuous variables. An example of the other kind of LVM is the mixture distribution in which the latent variable is the discrete component label, and the corresponding models are called mixture LVMs, which will also be discussed in Section 3.2.

### 3.2. Theory and formulation of popular LVMs

Aiming at providing a more thorough tutorial on traditional LVMs, we discuss the theory and formulation of some typical LVMs in this section. We start from one of the simplest LVMs which has good probability interpretation — FA.

#### 3.2.1. Factor analysis

FA supposes that the observed variable $x$ is generated from the latent variable $t$ in the following way:

$$x = Wt + \mu + \epsilon \tag{4}$$

where $W \in \mathbb{R}^{m \times k}$ is the mapping matrix that maps $t$ to $x$, also known as factor loading matrix, and $\mu$ is a bias vector. $\epsilon$ is an additive noise term whose distribution is a Gaussian with zero-mean and covariance $\Psi = \text{diag}(\psi_1, \psi_2, \ldots, \psi_m)$.

The objective of FA is to estimate the unknown model parameters in (4), especially $W$ and $\Psi$. However, it is impossible to compute $W$ and $\Psi$ unless we make some further assumptions about (4). As mentioned, a crucial assumption in FA is that the prior distribution of latent variable $t$ is a zero-mean unit covariance Gaussian distribution $\mathcal{N}(0, I)$. This will yield, according to (4), a Gaussian as the marginal distribution of $x$:

$$p(x) = \mathcal{N}(\mu, WW^T + \Psi). \tag{5}$$

Generally, the purpose of FA is to estimate the parameters $\{\mu, W, \Psi\}$ using the training dataset $X \in \mathbb{R}^{n \times m}$ (n is the number of samples). Through maximum likelihood, it is easy to obtain that $\mu$ is equal to dataset mean $\bar{x}$, i.e. $\mu = \bar{x} = \frac{1}{n}\sum_{i=1}^{n} x_i$. The parameters $W$ and $\Psi$ can be efficiently estimated through the expectation–maximization (EM) algorithm (Bishop, 2007). Specifically, in the E-step, we compute the following equations:

$$\mathbb{E}[t_i] = GW^T\Psi^{-1}(x_n - \bar{x}) \tag{6}$$

$$\mathbb{E}[t_i t_i^T] = G + \mathbb{E}[t_i]\mathbb{E}[t_i]^T \tag{7}$$

where $G$ is calculated by

$$G = (I + W^T\Psi^{-1}W)^{-1}. \tag{8}$$

Then in the M-step, the updates of $W$ and $\Psi$ are given by

$$W_{\text{new}} = \left[\sum_{i=1}^{n}(x_i - \bar{x})\mathbb{E}[t_i]^T\right]\left[\sum_{i=1}^{n}\mathbb{E}[t_i t_i^T]\right]^{-1} \tag{9}$$

---

[3] In some cases, the *effects* of underlying generators on observed variables are also treated as the latent variables. For example, in GMM, the discrete component label — which measures how the observed sample is generated from the mixture components — is regarded as the latent variable.

 *Annual Reviews in Control xxx (xxxx) xxx*

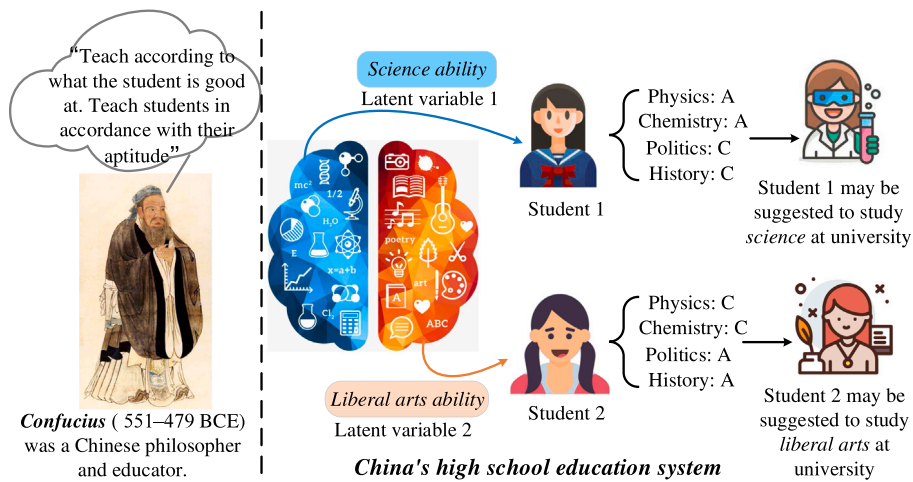

"Teach according to what the student is good at. Teach students in accordance with their aptitude"

***Confucius*** ( 551–479 BCE) was a Chinese philosopher and educator.

**Fig. 3. A vivid illustration of what is a latent variable.** Confucius, a famous ancient Chinese philosopher and educator, once put forward the idea of "teach students in accordance with their aptitude", which has influenced China for thousands of years until today. From 1977 to 2014, China's high school education system and college entrance examination both followed a manner called "dividing liberal arts and sciences", which means schools and teachers may suggest some students study only science subjects and some students study only liberal arts subjects. So how to distinguish whether a student should study science or liberal arts? Naturally, the teacher may recommend people with higher aptitude or ability in science to study scientific subjects, and people with higher ability in liberal arts to study liberal arts. However, there is no way to directly observe the science ability or liberal arts ability. We cannot straightforwardly obtain a specific value of a student's science ability. But we know that a student with high ability in science is more likely to get higher test scores in scientific subjects. In other words, although a student's science or liberal arts ability cannot be directly observed, it may serve as a fundamental factor in affecting the student's test scores. For example, if a student gets both A in physics and chemistry but only C in politics and history, the teacher is more likely to recommend him/her to study science. Conversely, a student with high test scores in politics and history and low scores in physics and chemistry may be more suitable for studying liberal arts. In this case, the students' ability in science or liberal arts cannot be observed, but the test scores of the corresponding subjects can be directly seen. We will infer which aspect of the student's ability is higher according to the student's test scores, and then recommend him/her to study science or liberal arts. This is a typical example of how latent variables serve as the underlying generators to affect observed variables.

$$\Psi_{\text{new}} = \text{diag}\left(S - W_{\text{new}} \frac{1}{n} \sum_{i=1}^{n} \mathbb{E}[t_i](x_i - \bar{x})^T\right) \tag{10}$$

where the $\text{diag}(\cdot)$ sets a matrix's all non-diagonal elements to zero. $S$ is the data covariance matrix computed by

$$S = \frac{1}{n} \sum_{i=1}^{n} (x_i - \bar{x})(x_i - \bar{x})^T \tag{11}$$

So far, the construction of E-step and M-step is complete.

On the other hand, in the FA model, if the covariance $\Psi$ is isotropic rather than diagonal, i.e. all $\psi_j (j = 1, \ldots, m)$ values are assumed to be the same, then FA is equivalent to the probabilistic PCA (PPCA). Lawley (Lawley, 1953) and Anderson (Anderson, 1963) first discovered the connections between PCA and FA. They proved that for a FA model with $\Psi = \sigma^2 I$, the stationary points of the likelihood function will yield a factor loading matrix $W$ whose columns are eigenvectors of the sample covariance matrix, and $\sigma^2$ is the mean of the non-principle eigenvalues of PCA. Later, Tipping and Bishop (Tipping & Bishop, 1999) further studied such a relationship and proposed the PPCA model. PPCA can also be solved through the EM algorithm (Bishop, 2007; Ge, 2018), which shares similar procedures as FA, so we do not discuss it in detail. Instead, this paper focuses on the theory of the original PCA model.

### 3.2.2. Principal component analysis

The Hotelling's definition (Hotelling, 1933) is used to explain PCA here. Suppose the training dataset $X \in \mathbb{R}^{n \times m}$ is mean-centered and variance-scaled. PCA pursues a projection vector $p$ that maximizes the variance of the projected data $t = Xp$, which can be formulated as the following optimization task:

$$\arg\max_p \frac{1}{n} t^T t = \arg\max_p \frac{1}{n} p^T X^T X p$$

$$\text{s.t. } p^T p = 1 \tag{12}$$

where the constraint is introduced to prevent $\|p\| \to \infty$. And since we only focus on the direction of $p$, not its magnitude, a simple and reasonable way is to set $p$ as a unit vector.

Using Lagrange multiplier approach to solve task (12), we can obtain the following eigenvalue decomposition problem:

$$\frac{1}{n} X^T X p = \lambda p. \tag{13}$$

The solution of (13) yields $m$ eigenvalues and their corresponding eigenvectors. Denote $\lambda_1, \lambda_2, \ldots, \lambda_m$ as the eigenvalues which are sorted in descending order. The eigenvector $p_j \in \mathbb{R}^m$ of $\lambda_j$ ($1 \leqslant j \leqslant m$) is also called loading vector. The component $t_j \in \mathbb{R}^N$, which is obtained by the linear transformation $t_j = Xp_j$, is also known as the score vector.

PCA divides the loading matrix $P = [p_1, p_2, \ldots, p_m] \in \mathbb{R}^{m \times m}$ into two submatrices. While the first $k$ vectors $P_p = [p_1, p_2, \ldots, p_k] \in \mathbb{R}^{m \times k}$ construct the principal component subspace, and the rest $(m-k)$ vectors $P_r = [p_{k+1}, p_{k+2}, \ldots, p_m] \in \mathbb{R}^{m \times (m-k)}$ represent the residual subspace. Through $P_p$ and $P_r$, the whole score matrix can be represented by

$$T = [T_p, T_r] = X[P_p, P_r]. \tag{14}$$

The submatrix $T_p$ is the features extracted by PCA, whose columns are also known as principal components (PCs). PCA assumes that the observed dataset $X$ is generated by the PCs, and the mapping matrix is $P_p^T$:

$$X = T_p P_p^T + E \tag{15}$$

where $E \in \mathbb{R}^{n \times m}$ is called residual or error matrix. As can be seen, $T_p$ in (15) are the unobservable factors that form the manifest variables $X$, which is consistent with Definition 1 and the latent variables in PCA are just the extracted PCs.

Similar to FA, previous studies (Bishop, 2007; Goodfellow, Bengio, & Courville, 2016) have shown that the latent variables in PCA also follow the Gaussian distribution. Based on the nice properties of the Gaussian distribution, Gaussian LVMs are relatively easy to analyze and fit data, thus having great practical significance. However, one should also note the real-world application scenarios are various, some of which may not be properly modeled with strict Gaussian latent variables. The consideration of non-Gaussian latent variable distributions is natural and necessary, which brings us the ICA model.

### 3.2.3. Independent component analysis

ICA is a statistical method for transforming an observed multi-dimensional random vector into components that are statistically as independent from each other as possible. It was first introduced to deal with the blind signal separation problem. The fundamental restriction in ICA is that the independent components must be non-Gaussian.

ICA is a generative model, it assumes an underlying process that generates the observed data. Given an original dataset $X \in \mathbb{R}^{m \times n}$ with $m$ variables and $n$ samples (note that the $X$ here is the transpose of that in PCA), ICA supposes the dataset $X$ can be represented as linear combinations of several independent sources, i.e.

$$X = AS \tag{16}$$

where $S \in \mathbb{R}^{k \times n}(k \le m)$ is the independent sources matrix, while $A \in \mathbb{R}^{m \times k}$ is the mixing matrix. The independent components $S$ are the latent variables in ICA, which are assumed to be the driving variables behind the whole system, and the purpose of ICA is to recover them.

Eq. (16) enlightens us if we invert $A$ and multiply it with the original dataset $X$, the independent sources will be retrieved. Define the separating matrix $W = A^+$, where $^+$ represents the pseudo inverse, then the estimation of source signals can be computed by

$$\hat{S} = WX. \tag{17}$$

The objective of ICA is to find a matrix $W$ which makes the components in $\hat{S}$ as independent of each other as possible.

The preprocessing steps of ICA not only require data normalization, a more important procedure is called sphering or whitening, which is designed to achieve the decorrelation of original data. Given a normalized matrix $X$ with its covariance matrix $E(XX^T)$, where $E(\cdot)$ stands for expectation. The eigenvalue decomposition of covariance matrix is calculated by

$$E(XX^T) = P\Lambda P^T. \tag{18}$$

Define $V = \Lambda^{-\frac{1}{2}} P^T$, the whitening transformation can be expressed as

$$Z = VX. \tag{19}$$

It is easy to verify that $E(ZZ^T) = I$.

Hyvärinen and Oja (2000) have proved that independent is equivalent to non-Gaussian, which provides great convenience for estimating $\hat{S}$. The most famous ICA algorithm is the FastICA (Hyvärinen & Oja, 2000), which is based on maximizing the non-Gaussianity of variables.

### 3.2.4. Canonical correlation analysis

FA, PCA and ICA are all models that extract latent variables from only a group of variables, but sometimes we need to model the relationship between several groups of variables. For example, the relationship between a student's height and weight and his/her test scores on long jump and high jump. Or the relationship between temperature and pressure of the chemical process and the concentration of several products. The consideration of the relationship between several variable groups yields some cross-modeling LVMs, in which a basic and typical model is the canonical correlation analysis (CCA).

Without loss of generality, we suppose there are two groups of data. CCA seeks two different subspaces for the first group and the second group respectively, so that the correlation between a pair of components in these two subspaces reaches the maximum. Let $X \in \mathbb{R}^{n \times m}$ and $Y \in \mathbb{R}^{n \times d}$ denote the first and second dataset respectively, where $n$ is the number of samples, $m$ and $d$ are the corresponding variable dimensions. Here $X$ and $Y$ are assumed to be centralized. Define two projection vectors $p$ and $q$, then the objective of CCA can be expressed as

$$\arg\max_{p,q} \ \text{corr}(Xp, Yq) = \frac{\text{cov}(Xp, Yq)}{\sqrt{D(Xp)}\sqrt{D(Yq)}} \tag{20}$$

where corr($\cdot$), cov($\cdot$), $D(\cdot)$ are the correlation, covariance, and variance operators respectively. Since $X$ and $Y$ are centralized matrices, we have

$$\text{cov}(Xp, Yq) = p^T X^T Yq$$
$$D(Xp) = p^T X^T Xp$$
$$D(Yq) = q^T Y^T Yq. \tag{21}$$

Therefore, (20) is equivalent to

$$\arg\max_{p,q} \ \frac{p^T X^T Yq}{\sqrt{p^T X^T Xp}\sqrt{q^T Y^T Yq}}. \tag{22}$$

Here we use a small trick, which is fixing the denominator to optimize the numerator, and (22) can be rewritten as

$$\arg\max_{p,q} \ p^T X^T Yq$$
$$\text{s.t. } p^T X^T Xp = 1, q^T Y^T Yq = 1 \tag{23}$$

Eq. (23) is an optimization task with equality constraints. By introducing two Lagrange multipliers $\lambda/2$ and $\mu/2$, we can obtain the following Lagrangian function

$$\mathcal{L} = p^T X^T Yq - \frac{\lambda}{2}(p^T X^T Xp - 1) - \frac{\mu}{2}(q^T Y^T Yq - 1). \tag{24}$$

Calculating the partial derivatives $\frac{\partial \mathcal{L}}{\partial p}$ and $\frac{\partial \mathcal{L}}{\partial q}$, and let them equal to 0, we have

$$\frac{\partial \mathcal{L}}{\partial p} = X^T Yq - \lambda X^T Xp = 0. \tag{25}$$

$$\frac{\partial \mathcal{L}}{\partial q} = Y^T Xp - \mu Y^T Yq = 0. \tag{26}$$

From (25) and (26), we can deduce that $\lambda = \mu = p^T X^T Yq$. That is to say, the Lagrange multiplier is the optimization objective, and we need to get a $\lambda$ as large as possible. Substituting (26) into (25), and (25) into (26), respectively, we can obtain the following equations:

$$(X^T X)^{-1} X^T Y(Y^T Y)^{-1} Y^T Xp = \lambda^2 p \tag{27}$$

$$(Y^T Y)^{-1} Y^T X(X^T X)^{-1} X^T Yq = \lambda^2 q \tag{28}$$

where if $(X^T X)^{-1}$ or $(Y^T Y)^{-1}$ is a singular matrix, the pseudo-inverse can be used.

Eqs. (27) and (28) are two eigenvalue decomposition problems, in which $\lambda^2$ is the largest eigenvalue, and $p$ and $q$ are the corresponding eigenvectors. Like PCA, if we continue to extract the 2nd, 3rd, $\ldots$, $k$th (where $k \le \min(m, d)$) largest eigenvalues and the corresponding eigenvectors, then the following two projection matrices can be formed:

$$P = [p_1, p_2, \ldots, p_k], Q = [q_1, q_2, \ldots, q_k] \tag{29}$$

Using $P$ and $Q$ to project $X$ and $Y$ respectively will yield two low-dimensional feature matrices:

$$U = XP, V = YQ \tag{30}$$

where $U$ and $V$ are the extracted latent variables by CCA, which generates the original dataset in the following way:

$$X = UP^T + E_X$$
$$Y = VQ^T + E_Y \tag{31}$$

where $E_X$ and $E_Y$ are the corresponding error matrices.

CCA has been widely used to analyze the correlation of multiple groups of data, another cross-modeling LVM which shares similar idea with CCA is the PLS method.

### 3.2.5. Partial least squares

As analyzed, CCA looks for a pair of projection directions that maximize the correlation of the projection of two groups of data. CCA only considers the correlation of the projected features, which may lead to the elimination of some important information of the original datasets $X$ and $Y$. If we want the projected features to be as correlated as possible, meanwhile carrying the most variations of original data just like PCA, then we arrive at the PLS model.

According to the aforesaid analysis, the projection vectors $p_1$ and $q_1$ of PLS can be designed as

$$\text{corr}(Xp_1, Yq_1) \to \max; D(Xp_1) \to \max; D(Yq_1) \to \max \tag{32}$$

Based on (32), the optimization objective can be chosen as

$$\text{corr}(Xp_1, Yq_1)\sqrt{D(Xp_1)}\sqrt{D(Yq_1)} = \text{cov}(Xp_1, Yq_1). \tag{33}$$

Introducing two constraints, the following optimization task is formulated:

$$\underset{p,q}{\arg\max}\ \text{cov}(Xp_1, Yq_1)$$
$$\text{s.t.} p_1^T p_1 = 1, q_1^T q_1 = 1 \tag{34}$$

where the constraints are added to prevent $\|p_1\|, \|q_1\| \to \infty$. And since we only focus on the direction of the projection vectors, not their magnitude, a simple and reasonable way is to set $p_1$ and $q_1$ as unit vectors.

The solving process of (34) is quite similar to (23), the corresponding Lagrangian function is

$$\mathcal{L} = p_1^T X^T Y q_1 - \frac{\lambda}{2}(p_1^T p_1 - 1) - \frac{\mu}{2}(q_1^T q_1 - 1). \tag{35}$$

Following similar steps as (25) and (26), the solution of (35) yields the following equations:

$$X^T Y Y^T X p_1 = \lambda^2 p_1 \tag{36}$$

$$Y^T X X^T Y q_1 = \lambda^2 q_1 \tag{37}$$

The computations of $p_1$ and $q_1$ in (36) and (37) are quite simple than those in (27) and (28). Eqs. (36) and (37) indicate $p_1$ and $q_1$ are the left and right singular vectors of $X^T Y$, respectively. Therefore, the projection vectors in PLS can be obtained through performing singular value decomposition (SVD) on $X^T Y$.

The first pair of latent variables $u_1$ and $v_1$ in PLS are calculated by $u_1 = Xp_1$ and $v_1 = Yq_1$. The computing of the subsequent pairs of latent variables in PLS is quite different with CCA. After obtaining $u_1$ and $v_1$, PLS performs rank-one deflation on $X$:

$$X = X - \frac{u_1 u_1^T}{\|u_1\|^2} X. \tag{38}$$

The new $X$ in (38) is again substituted into optimization problem (34), thus we solve the new solutions of (34) brings us the second pair of latent vectors $u_2$ and $v_2$. Repeat the above process $k$ times, we get two groups of latent variables $U = [u_1, u_2, \dots, u_k]$ and $V = [v_1, v_2, \dots, v_k]$. It is worth noting that, $k$ should be no greater than the rank of $X$. The above process shows that (34) can be directly solved through matrix decomposition. On the other hand, the latent variables of PLS can also be computed by iteration algorithm. The most popular iteration algorithm for PLS is the NIPALS algorithm (Wold, 1975), which computes the latent variables one by one through the power method (Wegelin, 2000).

The above modeling process does not specify the meaning of the matrices $X$ and $Y$. However, in most scenarios of PLS applications, $X$ is the observed variable matrix and $Y$ is the target variable matrix. In this situation, by performing matrix factorization on the cross-matrix $X^T Y$, PLS is essentially a method that models the association between observed variable matrix and target variable matrix by means of latent variables. The purpose of PLS is to predict $Y$ from $X$ with the help of latent variables. The continuous and discrete targets in $Y$ respectively yield the PLS-Regression and PLS-Classification (also known as PLS-Discriminant Analysis) models. PLS can be used for regression, classification, dimension reduction, feature extraction, etc. Among them, PLS-Regression is the most widely used method, for more details about it, please refer to Wold, Sjöström, and Eriksson (2001).

### 3.2.6. Gaussian mixture model

The aforesaid models are all continuous LVMs, in which the extracted latent factors are all continuous variables. There exists the other kind of LVM called discrete LVMs. A typical example of discrete LVM is the mixture model, in which the latent variable is the discrete component label. We start our analysis from the most popular mixture model — GMM.

Gaussian distribution is a widely used probability function in statistics, data mining, and machine learning. In many data mining scenarios, Gaussian distribution is the preferred probability function for sample fitting. However, when the distribution of real data is relatively complex, a single Gaussian distribution may not be able to effectively represent the data form. At this time, we may consider using a combination of multiple Gaussian distributions to fit the data, which yields the GMM. GMM can approximate complex data distribution by linearly combining several Gaussian distributions.

For a $m$-dimensional random vector $x$ that follows the Gaussian distribution, its probability density function can be expressed as

$$p(x; \mu, \Sigma) = \frac{1}{(2\pi)^{\frac{m}{2}}|\Sigma|^{\frac{1}{2}}} e^{-\frac{1}{2}(x-\mu)^T \Sigma^{-1}(x-\mu)} \tag{39}$$

where $\mu$ is the $m$-dimensional mean vector, $\Sigma \in \mathbb{R}^{m \times m}$ is the covariance matrix.

Based on (39), the GMM can be formed as

$$p(x) = \sum_{i=1}^{k} \alpha_i \cdot p(x; \mu_i, \Sigma_i) \tag{40}$$

where $\alpha_i \geq 0$ is the mixture coefficient and $\sum_{i=1}^{k} \alpha_i = 1$. $\mu_i$ and $\Sigma_i$ are the parameters of the $i$th Gaussian mixture component. The objective of GMM is to estimate the model parameters $\{(\alpha_i, \mu_i, \Sigma_i) | 1 \leq i \leq k\}$ in (40). There are many algorithms for solving GMM, such as the variational inference method, information entropy method and the numerical optimization method. Here we introduce the most commonly used EM algorithm.

Suppose there is a dataset $\{x_1, x_2, \dots, x_n\}$ in which $x_j \in \mathbb{R}^m (j = 1, 2, \dots, n)$. Define a variable $\gamma_{ji}$ that represents the $j$th sample $x_j$ is generated from the $i$th Gaussian component $p(x; \mu_i, \Sigma_i)$, which can be expressed as

$$\gamma_{ji} = \begin{cases} 1, & \text{if } x_j \text{ is generated from } p(x; \mu_i, \Sigma_i). \\ 0, & \text{otherwise.} \end{cases} \tag{41}$$

In GMM, the $k$-dimensional vector $\gamma_j = [\gamma_{j1}, \gamma_{j2}, \dots, \gamma_{jk}]$ is the latent variable, which measures how the observed sample $x_j$ is generated by the mixture Gaussian components. With the help of $\gamma_{ji}$, the solving procedures of EM algorithm for GMM can be formulated as follows.

1. Randomly initialize the model parameters $\{(\alpha_i, \mu_i, \Sigma_i) | 1 \leq i \leq k\}$.
2. In the E-step, computing the posterior probability of $\gamma_{ji}$:

$$\hat{\gamma}_{ji} = \frac{\alpha_i \cdot p(x_j; \mu_i, \Sigma_i)}{\sum_{i=1}^{k} \alpha_i \cdot p(x_j; \mu_i, \Sigma_i)} \tag{42}$$

3. In the M-step, update the model parameters:

$$\hat{\mu}_i = \frac{\sum_{j=1}^{n} \hat{\gamma}_{ji} x_j}{\sum_{j=1}^{n} \hat{\gamma}_{ji}}, i = 1, 2, \dots, k$$
$$\hat{\Sigma}_i = \frac{\sum_{j=1}^{n} \hat{\gamma}_{ji}(x_j - \hat{\mu}_i)(x_j - \hat{\mu}_i)^T}{\sum_{j=1}^{n} \hat{\gamma}_{ji}}, i = 1, 2, \dots, k$$
$$\hat{\alpha}_i = \frac{\sum_{j=1}^{n} \hat{\gamma}_{ji}}{n}, i = 1, 2, \dots, k \tag{43}$$

4. Repeat steps (2) and (3) until convergence.

Actually, by replacing $p(x; \mu_i, \Sigma_i)$ in (40) to other probability distributions, we can also get the mixed model of other distributions, such as the *Bernoulli mixture model*, which is also know as *latent class analysis* (McLachlan, Lee, & Rathnayake, 2019).

In continuous LVMs, the latent variables can be regarded as the extracted features which can be used for other learning tasks. While for GMM, the latent variable $\gamma_j$ usually cannot be treated as features, since it only measures how the observation is generated by the mixture components, but does not contain information about the components themselves. In GMM, the more important elements are the mixture components, which can be regarded as the actual underlying generators of the observation. Such a phenomenon corresponds to the footnote part of Definition 1.

### 3.2.7. Hidden Markov model

Another representative discrete LVM is the hidden Markov model (HMM), which belongs to the families of probabilistic graphical models. HMM describes a process in which an unobservable *state sequence* is randomly generated from a hidden Markov chain, and then the *observation sequence* is generated based on the state sequence.

The classic HMM is a discrete model, which assumes that the hidden states and the observations come from two different finite sets. Denote $V$ as the set of observations, and $Q$ as the set of hidden states:

$$V = \{v_1, v_2, \ldots, v_m\}, Q = \{q_1, q_2, \ldots, q_k\} \tag{44}$$

where $k$ is the number of possible states and $m$ is the number of possible observations.

Suppose there is an observation sequence $O$ with length $n$, and $S$ is the state sequence corresponding to $O$:

$$O = \{o_1, o_2, \ldots, o_n\}, S = \{s_1, s_2, \ldots, s_n\}. \tag{45}$$

In addition to these model structure information, the following three parameters are also required to determine an HMM:

1. The *transition probability matrix* $A = [a_{ij}]_{k \times k}$ that measures the probability of moving from one state to another:

   $$a_{ij} = P(s_{t+1} = q_j | s_t = q_i), \quad i, j = 1, 2, \ldots, k \tag{46}$$

   where $a_{ij}$ represents the probability that the state at time $t$ is $q_i$ and the state at the next time $t + 1$ is $q_j$.

2. The *emission probability matrix* $B = [b_{ij}]_{k \times m}$ that expresses the probability of an observation being generated from a state:

   $$b_{ij} = P(o_t = v_j | s_t = q_i), i = 1, 2, \ldots, k; j = 1, 2, \ldots, m \tag{47}$$

   where $b_{ij}$ represents the probability that at time $t$, the observation is $v_j$ when the state is $q_i$.

3. The *initial probability vector* $\pi = (\pi_1, \pi_2, \ldots, \pi_k)$:

   $$\pi_i = P(s_1 = q_i), \quad i = 1, 2, \ldots, k \tag{48}$$

   where $\pi_i$ represents the initial state of HMM is $q_i$.

The transition probability matrix $A$ and the initial state probability vector $\pi$ determine the Markov chain and generate an unobservable hidden state sequence. The emission probability matrix $B$ controls how to generate observations from the hidden states. A HMM can be obtained when $\pi$, $A$ and $B$ are determined. Therefore, we use $\lambda = (A, B, \pi)$ to represent a HMM. There are three basic problems in HMM (Rabiner, 1989):

1. **Likelihood:** Given an HMM $\lambda = (A, B, \pi)$ and an observation sequence $O$, determine the likelihood $P(O | \lambda)$. Efficient algorithms include the *forward–backward algorithm* (Rabiner, 1989).
2. **Decoding:** Given an observation sequence $O$ and an HMM $\lambda = (A, B, \pi)$, discover the best hidden state sequence $S$. Typical algorithms include the *Viterbi algorithm* (Viterbi, 1967).

3. **Learning:** Given an observation sequence $O$, learn the HMM parameters $\lambda = (A, B, \pi)$. Typical algorithms include the *Baum–Welch algorithm* (Baum, Petrie, Soules, & Weiss, 1970).

Among them, the decoding problem can be regarded as the process of extracting latent variables.

### 3.2.8. Extensions of classic LVMs

In the above discussion, we present five continuous LVMs and two discrete LVMs. Among them, the FA, PCA, ICA, GMM and HMM belong to unsupervised learning. The CCA and PLS are considered as two supervised learning methods. However, these classic LVMs, especially when they are in the probabilistic framework, can be extended to other more complex forms through some techniques.

By incorporating labeled data into the modeling process, unsupervised models such as FA and PCA can be extended to supervised forms (Barshan, Ghodsi, Azimifar, & Jahromi, 2011; Yang, Yao, & Ge, 2020; Yao & Ge, 2017b; Yu, Yu, Tresp, Kriegel, & Wu, 2006). By incorporating labeled and unlabeled data into the modeling process, supervised models such as PLS can be extended to semi-supervised forms (Bao, Yuan, & Ge, 2015; Li, Liu, & Huang, 2020; Yao & Ge, 2016; Zheng & Song, 2018). As can be seen, most traditional LVMs only involve linear transformations. In the real world applications, especially the industrial process, the systems may have nonlinear properties. In order to cope with such situations, kernel trick is introduced to extend classic LVMs to nonlinear forms, such as kernel PCA (Deng, Tian, & Chen, 2013; Deng & Wang, 2018; Fan, Chow, & Qin, 2021; Ge, Yang, & Song, 2009; Lee, Yoo, Choi, Vanrolleghem, & Lee, 2004; Lee, Yoo, & Lee, 2004a; Schölkopf, Smola, & Müller, 1997; Yuan, Ge, & Song, 2014), kernel ICA (Bach & Jordan, 2002; Cai, Tian, & Chen, 2015; Jianwei, Ye, He, & Jin, 2021; Lee, Qin, & Lee, 2007; Tian, Zhang, Deng, & Chen, 2009), kernel PLS (Peng, Zhang, & Li, 2013; Rosipal & Trejo, 2001; Si, Wang, & Zhou, 2020; Zhang, Zhou, Qin, & Chai, 2009; Zhu, Zhao, & Liu, 2021). By integrating temporal information in the modeling process, the dynamical variants of classic LVMs can be obtained (Fan & Wang, 2014; Huang, Yi, & Li, 2020; Ku, Storer, & Georgakis, 1995; Lee, Yoo, & Lee, 2004b; Peterson, Wilberg, Cortés, & Latour, 2021; Prawin & Rao, 2018; Shi, Guo, et al., 2020; Wu, Chan, Tsui, & Hou, 2013; Yu, et al., 2021; Zuur, Fryer, Jolliffe, Dekker, & Beukema, 2003).

Aiming at solving different kind of problems, the extensions and variants make improvements for classic LVMs in various aspects, which gives them more powerful representational and modeling ability, and further expand their application scopes.

### 3.3. Industrial applications of typical LVMs

Classic LVMs and their variants play an important role in many industrial scenarios. Among various applications, the two most popular ones are *process monitoring* (Ge, Song, & Gao, 2013; Wang, Si, Huang, & Lou, 2018) and *soft sensor* (Jiang, Yin, Dong, & Kaynak, 2020; Kadlec, Grbić, & Gabrys, 2011). This paper will focus most of the attention on applications in these two areas, and will briefly sort out some applications in other fields.

Process monitoring mainly includes *fault detection* and *fault diagnosis*. Fault detection is a kind of technique that judges *whether the process is under fault condition or not*, or *if an abnormal event happens in the process or not*. Fault diagnosis refers to *identifying the type of the detected faults and locating which part or component of the process is abnormal*, or *providing the root cause of the detected fault.*[4]

---

[4] Under finer grained division, process monitoring may include fault detection, fault diagnosis, fault identification and fault isolation. In some literature, the latter three problems are also collectively called fault diagnosis, and this paper adopts this division approach.

Soft sensor is also called *soft sensing*, *virtual sensor*, or *virtual sensing*, which refers to *methods and algorithms that are used to estimate or predict certain physical quantities or product quality in industrial processes based on accessible measurements, information and knowledge.* (Jiang et al., 2020) The difference between soft sensors and physical sensors is that their measurement principle is driven by data and knowledge, and meanwhile they are implemented on computer software-based systems or embedded systems. To sum up, soft sensor is a kind of technique that uses easy-measured auxiliary variables to predict hard-measured quality variables. Soft sensors can deal with the obstacle that some physical quantities cannot be measured timely and accurately.

Generally, soft sensor can be regarded as a regression problem. The first part of fault diagnosis — identifying the type of the detected faults — can be seen as fault identification, which is essentially a classification problem. Fault detection and the second part of fault diagnosis — locating which component of the process is abnormal — are two unsupervised issues, whose solving processes involve *constructing monitoring statistic* and *estimating the contribution of the variable to the fault* (Qin, 2012).

### 3.3.1. Application status

This subsection generally discusses the application status of classic LVMs in some typical industrial areas. First of all, the overall application status of classic LVMs in the whole industrial modeling field is shown in Fig. 4. Fig. 4(a) uses the pie chart to illustrate the application status of classic LVMs in industrial modeling. As can be seen, the most popular LVMs in the industry community are PCA and FA. Figs. 4(b) and 4(c) display the publishing trends of PCA and FA in the industrial modeling area, respectively. Publications using PCA and FA both have obvious growth after 2014.

Next, we discuss the applications of LVMs in several specific industrial areas. The application status in fault detection and fault diagnosis are shown in Figs. 5 and 6, respectively. It can be seen that PCA plays a dominant role in the applications of these two fields. The second popular method in fault detection and fault diagnosis is the ICA model, which occupies 12% and 13% of the share, respectively. The publishing trends of PCA and ICA in fault detection are shown in Figs. 5(b) and 5(c), and their publishing trends in fault diagnosis are plotted in Figs. 6(b) and 6(c). Another popular industrial application field of LVMs is the soft sensor area, whose application statistics are drawn in Fig. 7. Since soft sensor is essentially a regression problem, the regression oriented PLS model is quite suitable for it, thus making PLS plays a leading role in this field.

It can be seen that no matter in Fig. 4, Fig. 5, Fig. 6, or Fig. 7, there are far more publications from 2014 to 2021 than in previous years, which indicates that the applications of LVMs in industrial modeling has maintained a high degree of activity recently and attracted lots of attention in both academia and industry communities. Fig. 8 illustrates the top eight industrial application areas of classic LVMs. The most widely used area of LVMs is the chemical field, including Engineering Chemical, Chemistry Analytical and Chemistry Applied, which accounts for the largest proportion at about 30%. Other popular application areas include Environmental Sciences (21%), Food Science Technology (17%), etc.

## 4. Deep LVMs

This section first presents a thorough introduction to current mainstream DLVMs with emphasis on their theories and model architectures, then provides a detailed survey on industrial applications of mainstream DLVMs.

### 4.1. An introduction to mainstream DLVMs

Mainstream DLVMs mainly include DBN, AE and its variants, which will all be elaborated in the following subsections.

#### 4.1.1. Deep belief network

A DBN is a deep structure made by stacked restricted Boltzmann machines (RBMs). It is first proposed by Hinton et al. (2006) to solve the "explaining away" problem and fasten the estimation process of conditional distribution. To clarify DBN, RBM will be firstly introduced.

RBM is a Markov random field with hidden neurons. As Fig. 9 indicates, it consists of two layers: a hidden layer and a visible layer. To simplify the calculation of posterior, RBM assumes there are no intra-layer connections, which means all the stochastic neurons in the same layer are independent. At the same time, the hidden and visible layers are fully connected with undirected edges. For simplicity, neurons are assumed to be stochastic binary variables.

In Fig. 9, the hidden layer is represented as $h = [h_1, h_2, \dots, h_m]^T \in \{0,1\}^m$, and the visible layer is $v = [v_1, v_2, \dots, h_n]^T \in \{0,1\}^n$. $\alpha = [\alpha_1, \alpha_2, \dots, \alpha_m]^T \in \mathbb{R}^m$ and $\beta = [\beta_1, \beta_2, \dots, \beta_n]^T \in \mathbb{R}^n$ are column bias vectors for each layer, while $W \in \mathbb{R}^{m \times n}$ symbolizes the weight matrix.

Since RBM is a Markov random field, the joint probability is:

$$p(v, h) = \frac{1}{Z} \exp(-E(v, h)) \tag{49}$$

where energy function:

$$E(v, h) = -\beta^T v - \alpha^T h - h^T W v \tag{50}$$

and partition function:

$$Z = \sum_{h,v} exp(-E(v, h)) \tag{51}$$

Because of the intra-layer independence, the marginal distribution of RBM can be expressed as:

$$p(h_i = 1|v) = \sigma(\alpha_i + \sum_{j=1}^{n} w_{i,j} v_j) \tag{52}$$

$$p(v_i = 1|h) = \sigma(\beta_i + \sum_{j=1}^{m} w_{j,i} h_j) \tag{53}$$

where $\sigma(\cdot)$ denotes the logistic sigmoid function.

Define parameters $\theta = \{W, \alpha, \beta\}$, the log-likelihood function of $\theta$ on the dataset $V$ with size $N$ is computed as:

$$\mathcal{L}(\theta) = \frac{1}{N} \sum_{\hat{v} \in V} \log(p(\hat{v}))$$

$$= \mathbb{E}_{data}[\log(p(\hat{v}))] \tag{54}$$

The learning process of RBM is to find parameters $\theta$ that maximize the log-likelihood function such that the model can effectively describe the dataset. It can be written as:

$$\hat{\theta} = \arg\max_{\theta} \mathbb{E}_{data}[\log(p(\hat{v}))]$$

$$= \arg\min_{\theta} \text{KL}(P_{data}(\hat{v}) \parallel P_{model}(\hat{v})) \tag{55}$$

where $\text{KL}(\cdot)$ is the Kullback–Leibler (KL) divergence. Hinton et al. (Hinton et al., 2006) provide a contrastive divergence (CD) algorithm to update parameters, which is based on $k$-step blocked Gibbs sampling.

We can clamp the visible layer of one RBM (change undirected edges to directed edges), and regard the hidden layer as a visible layer of another RBM. Repeating this procedure, DBN can be established by stacked RBMs. A typical DBN consists of multiple RBMs, which is illustrated in Fig. 10. Once the visible layer is clamped, the likelihood function $p(v|h)$ is fixed, while the posterior probability $p(h|v)$ can be trained. The bottom layer of DBN is a visible vector that receives input data. This Multilayer structure enables DBN to extract complicated nonlinear features.

As a generative model, DBN uses hidden layers to extract the potential probability distribution of data. All the hidden neurons form a hidden space to represent abstract features. Using a simple up–down algorithm, DBN can generate new data according to these features.

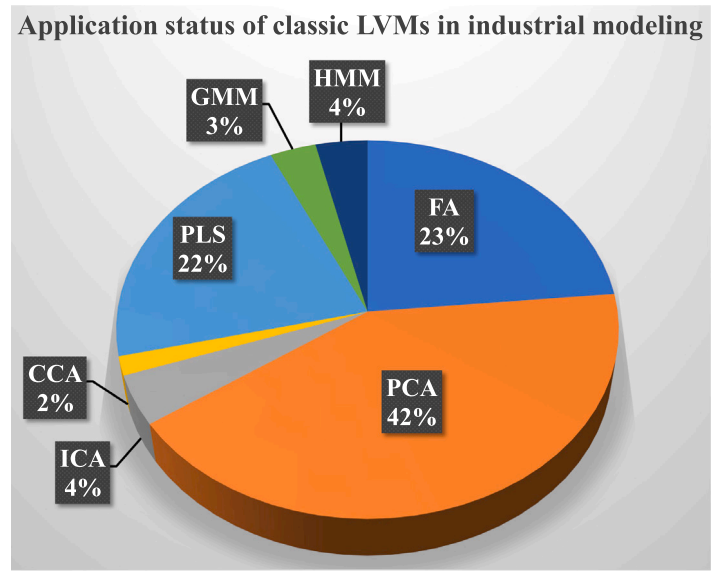

(a) General application status of classic LVMs in industrial modeling.

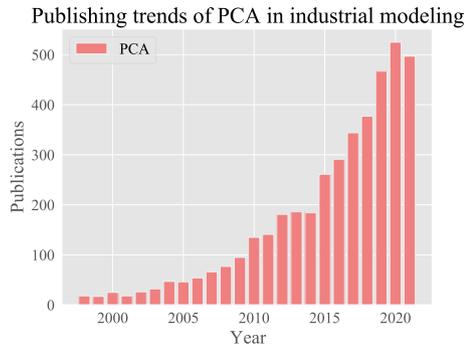

(b) Publishing trends of PCA in industrial modeling (1998-2021).

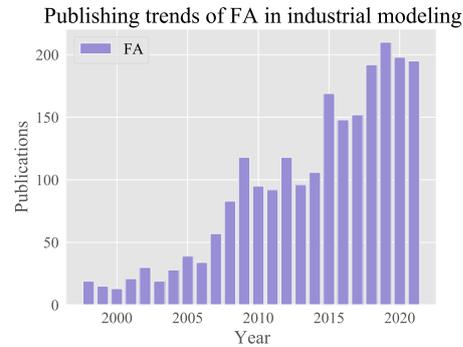

(c) Publishing trends of FA in industrial modeling (1998-2021).

**Fig. 4.** Some application statistics of classic LVMs in industrial modeling.

DBN's training process is divided into two different parts: unsupervised pre-training and supervised discriminative fine-tuning. The pre-training procedure adopts a greedy layer-by-layer algorithm to train RBMs. The algorithm regards every adjacent two layers of DBN as a traditional RBM and uses the CD algorithm to update weights separately. Previous RBM yields conditional probability $p(h|v)$ to sample the hidden layer, and then these samples are passed to the upper layer as input data. The complexity is linearly related to the depth and size of the network (Pandey & Janghel, 2019).

The pre-training procedure focuses on the inner relationship among data and builds a latent space to represent abstract features. It serves as the initialization of the entire network. For classification tasks, the network also demands a classifier such as BP networks to map abstract features onto labels. The classifier can be trained by labeled data and fine-tunes the DBN from top to bottom. The algorithm that updates the pre-trained weights in DBN is called the "wake–sleep" algorithm, which is a fast and economical method to learn representations (Hinton, Dayan, Frey, & Neal, 1995). The supervised fine-tuning process promotes DBN to find the global optimum.

The combination of pre-training and fine-tuning prevents the network from being trapped in local optimums. It also fastens the data processing rate, which improves DBN's performance in big data situations.

### 4.1.2. Autoencoder

The autoencoder (AE) originated from the research on Boltzmann machines. Ackley et al. developed an algorithm for Boltzmann machines and proposed "the encoder problem" to test the learning algorithm (Ackley, Hinton, & Sejnowski, 1985). Rumelhart et al. firstly constructed a model whose input is mapped into output through a small hidden layer, which is the embryo of AE (Rumelhart, Hinton, & Williams, 1986). Later, Bourlard et al. explained AE's structure and mathematical principles in detail (Bourlard & Kamp, 1988). Hinton and Zemel (1993) provided a new target function based on the minimum description principle to discover non-linear, factorial representations.

As a generative model, AE has been widely applied in unsupervised and weakly supervised learning. It accepts unlabeled samples and extracts abstract internal representations. Unlike traditional artificial neural networks, AE is supposed to reproduce original samples in the output layer. Therefore, a conventional AE is composed of an encoder and a decoder, like Fig. 11.

In the encoder, the model receives input data $x$ to extract internal representations $h$ in the hidden layer with weight matrix $W_e$ and bias $b_e$:

$$h = f(W_e x + b_e) \tag{56}$$

The decoder part inverses the process of encoder and tries to reproduce samples with weight $W_d$ and bias $b_d$:

$$\hat{x} = g(W_d h + b_d) \tag{57}$$

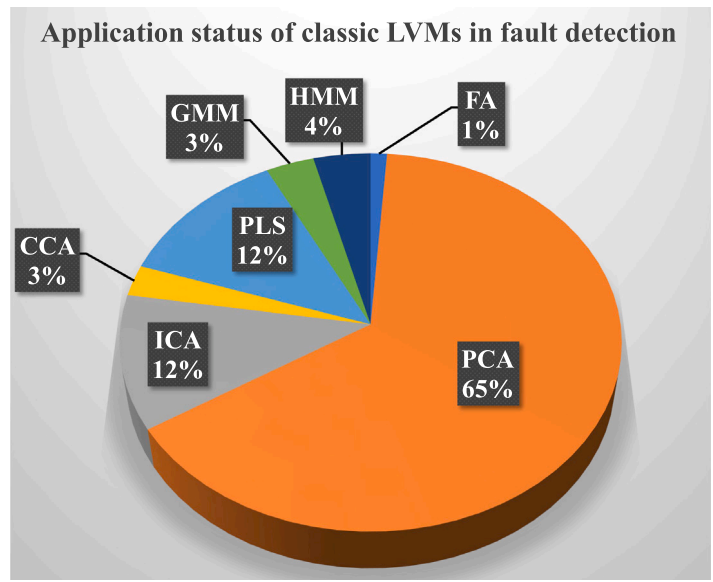

(a) General application status of classic LVMs in fault detection.

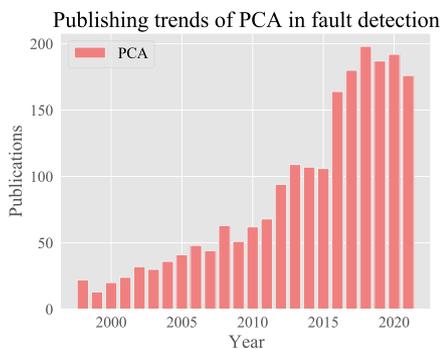

(b) Publishing trends of PCA in fault detection (1998-2021).

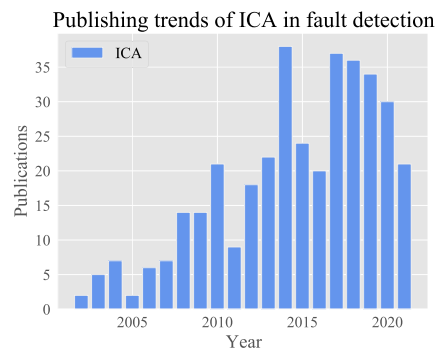

(c) Publishing trends of ICA in fault detection (2002-2021).

**Fig. 5.** Some application statistics of classic LVMs in fault detection.

where function $f(\cdot)$ and $g(\cdot)$ denote activation functions that determine whether the model is linear or non-linear. The training of AE is to minimize the loss function $\mathcal{L}$ that is usually chosen as either squared error loss or cross-entropy loss:

$$\mathcal{L}_s = \|\boldsymbol{x} - \hat{\boldsymbol{x}}\|_2^2 \tag{58}$$

$$\mathcal{L}_c = -\sum_{i=1}^{n} [x_i \log \hat{x}_i + (1 - x_i) \log(1 - \hat{x}_i)] \tag{59}$$

With a loss function, the backpropagation algorithm is used to train AEs. A common technique named "tied weights" can save training time (i.e. $\boldsymbol{W}_d = \boldsymbol{W}_e^T$). This assumption eliminates half of the parameters and keeps the ability to restore information. It is reasonable in the primitive linear applications of AEs since $\boldsymbol{W}_e^T$ is close to the right inverse of $\boldsymbol{W}_e$ when $\boldsymbol{W}_e$ approaches an orthogonal matrix. However, research shows that even in a non-linear and complex situation, such as a deep non-linear AE, the "tied weights" technique yields similar results to "untied weights" while the former costs less time (Vincent et al., 2010). Notably, using both "tied weights" and (58), AE is the same as a traditional PCA (Hinton & Zemel, 1993).

AE and RBM share similar logic. The only difference is that AE adopts a discriminative way while RBM uses a stochastic approach. Thus, like the evolution from RBM to DBN, AEs also can be extended to a deep architecture. Hinton and Salakhutdinov (2006) proposed the deep AE structure to improve the ability of extracting complex nonlinear features. Deep AE is still a symmetric structure but has multiple hidden layers, as shown in Fig. 12.

Another deep structure of AE is called stacked AE (SAE) (Bengio, Lamblin, Popovici, & Larochelle, 2006; Larochelle, Erhan, Courville, Bergstra, & Bengio, 2007; Shin, Orton, Collins, Doran, & Leach, 2013), which shares the same structure with the original deep AE but has a different pre-training approach. Specifically, instead of treating each layer pair as a RBM, SAE algorithm adds a fake layer to the pair such that they can form a fake classical AE. Then, these fake classical AEs can be trained layer-by-layer and the original two layers will be retained. The input data is fed to the first fake AE and values in hidden layer serve as input data of next fake AE. The model architecture and learning process of SAE are shown in Fig. 13.

*4.1.3. Denoising autoencoder*

AE and deep AE can extract intrinsic features of input data via unsupervised approaches. However, a complex structure cannot avoid the over-fitting problem, mainly when data contains noises. To improve the robustness of AE, Vincent et al. proposed the denoising AE (DAE) (Vincent, Larochelle, Bengio, & Manzagol, 2008).

As shown in Fig. 14, DAE is a classical AE with a random corruption layer. The corruption layer serves as a mask that randomly forces $vn$ components of input data $\boldsymbol{x}$ to 0s, where $n$ is the number of features and $v$ is the given proportion of corruption. Assuming the data after corruption is $\bar{\boldsymbol{x}}$ and the output is $\hat{\boldsymbol{x}}$, the target is to find the best parameters such that $\hat{\boldsymbol{x}}$ is as close as possible to $\boldsymbol{x}$, although $\hat{\boldsymbol{x}}$ actually comes from $\bar{\boldsymbol{x}}$ instead of $\boldsymbol{x}$. In Vincent et al. (2008), Vincent et al. give an intuitive explanation to justify the process: humans can

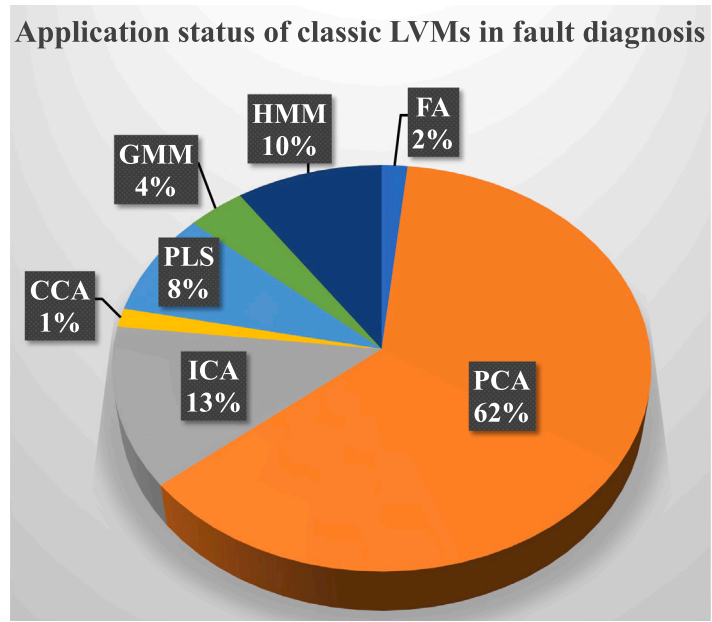

(a) General application status of classic LVMs in fault diagnosis.

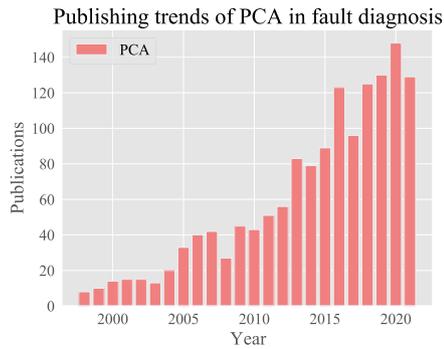
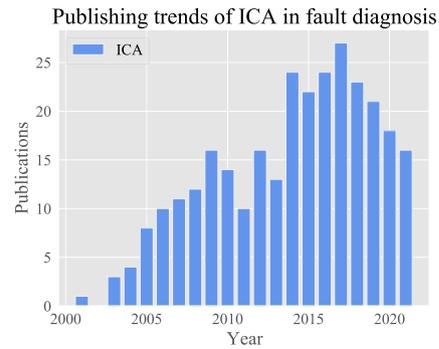

(b) Publishing trends of PCA in fault diagnosis (1998-2021).

(c) Publishing trends of ICA in fault diagnosis (2001-2021).

**Fig. 6.** Some application statistics of classic LVMs in fault diagnosis.

recognize partially corrupted images because of "the high-level concept associated to multiple modalities".

DAE has two advantages: it partially removes noises and narrows the gap between training and testing data. Thus, it can partially avoid the influence of noises and have better outcomes.

Vincent et al. combined DAE with SAE to build a deep architecture called stacked denoising AE (SDAE) (Vincent et al., 2010). The pre-training and fine-tuning processes are similar to SAE, but DAE is used as basic units instead of classical AE. Notably, the corruption layers in DAE and SDAE only work in the training process and are deactivated when predicting.

#### 4.1.4. Variational autoencoder

DAE corrupts some components of training data to solve the over-fitting problem. However, it is also possible to add Gaussian noises to the original data such that AEs can recognize unfamiliar inputs.

To achieve this goal, Kingma and Welling created a new architecture called variational AE (VAE) (Kingma & Welling, 2014). As shown in Fig. 15, $x = [x_1, x_2, \dots, x_n] \in \mathbb{R}^n$ is an input data from the training dataset, and it passes to the encoder network to learn the mean vector $\mu = [\mu_1, \mu_2, \dots, \mu_m] \in \mathbb{R}^m$ and standard deviation vector $\sigma = [\sigma_1, \sigma_2, \dots, \sigma_m] \in \mathbb{R}^m$ of latent variable $z = [z_1, z_2, \dots, z_m] \in \mathbb{R}^m$. $z$ will be decoded to regenerate $\hat{x} = [\hat{x}_1, \hat{x}_2, \dots, \hat{x}_n] \in \mathbb{R}^n$.

As a generative model, the best solution is to obtain $p(x)$, but it is impossible to calculate it directly. An alternative approach is assuming a latent variable $z$ such that:

$$p(x) = \int_z p(x|z)p(z)\,dz \tag{60}$$

VAE desires $p(z)$ to be a standard Gaussian distribution $z \sim \mathcal{N}(\mathbf{0}, \mathbf{I})$ and $p(x|z)$ to be a Gaussian distribution $x|z \sim \mathcal{N}(\mu_x(z), \sigma_x^2(z)\mathbf{I})$. Therefore, $p(x)$ is a consecutive Gaussian mixture model which can be easily calculated from $p(z)$ and $p(x|z)$ in a decoder network.

The design of VAE aims at achieving assumptions of $p(z)$ and $p(x|z)$. While the decoder part computes $\mu_x(z)$ and $\sigma_x(z)$ to regenerate an output $\hat{x}$, the encoder part tries to sample a latent vector $z$. However, for each individual sample, the encoder can only simulate the posterior probability $p(z|x)$. Since Gaussian distribution belongs to exponential family of distributions, $p(z|x) \propto p(z)p(x|z)$ is a Gaussian distribution $z|x \sim \mathcal{N}(\mu_z(x), \sigma_z^2(x)\mathbf{I})$. The encoder simulates $\mu_z(x)$ and $\sigma_z(x)$ to sample $z$. Actually, it calculates $\log(\sigma_z(x)^2)$ such that an activation function is not required to extend the symbol. However, the probability that the encoder gives is not the actual $p(z|x)$, so it is represented as $\tilde{p}(z|x)$.

The loss function $\mathcal{L}$ is defined as:

$$\mathcal{L} = \sum_{x \in Dataset} \log p(x) \tag{61}$$

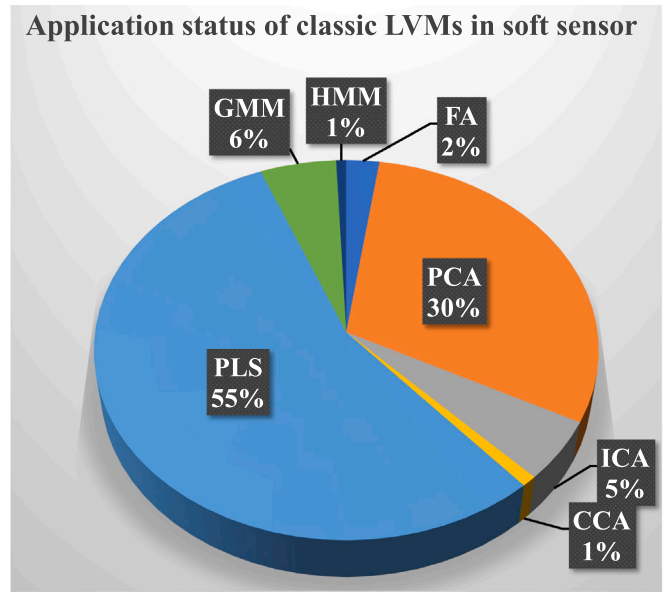

(a) General application status of classic LVMs in soft sensor.

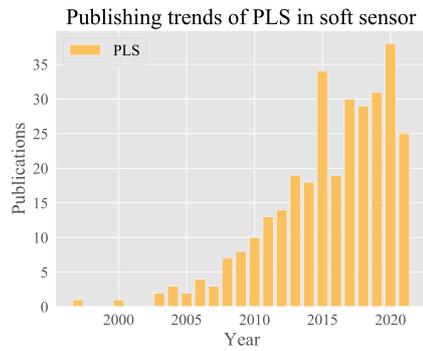

(b) Publishing trends of PLS in soft sensor (1997-2021).

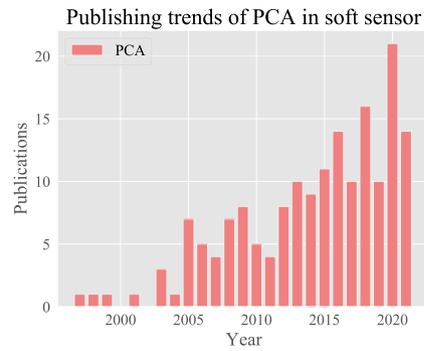

(c) Publishing trends of PCA in soft sensor (1997-2021).

**Fig. 7.** Some application statistics of classic LVMs in soft sensor.

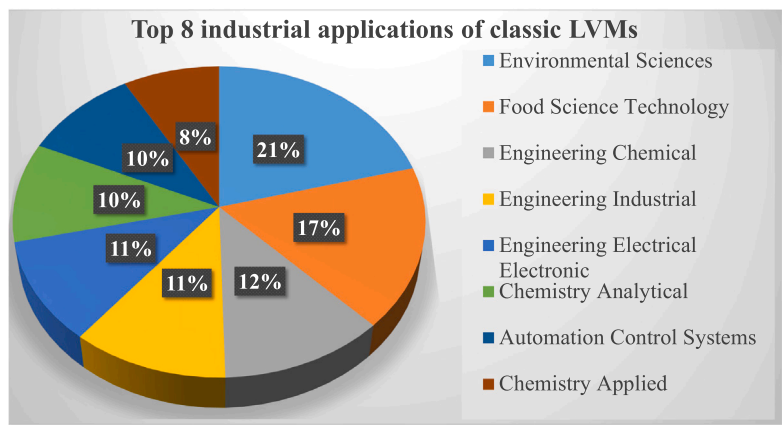

**Fig. 8.** Top eight industrial application areas of classic LVMs.

where $\log p(\boldsymbol{x})$ can be rewritten as:

$$\log p(\boldsymbol{x}) = \int_z \tilde{p}(z|\boldsymbol{x}) \log p(\boldsymbol{x}) \, dz$$

$$= \int_z \tilde{p}(z|\boldsymbol{x}) \log \frac{p(z, \boldsymbol{x})}{p(z|\boldsymbol{x})} \, dz$$

$$= \int_z \tilde{p}(z|\boldsymbol{x}) \log\left(\frac{p(z, \boldsymbol{x})}{\tilde{p}(z|\boldsymbol{x})} \frac{\tilde{p}(z|\boldsymbol{x})}{p(z|\boldsymbol{x})}\right) dz$$

$$= \int_z \tilde{p}(z|\boldsymbol{x}) \log \frac{p(z, \boldsymbol{x})}{\tilde{p}(z|\boldsymbol{x})} \, dz + KL(\tilde{p}(z|\boldsymbol{x}) \parallel p(z|\boldsymbol{x})) \tag{62}$$

The second term in (62) is the non-negative KL divergence between $\tilde{p}(z|\boldsymbol{x})$ and $p(z|\boldsymbol{x})$, while the first is the evidence lower bound (ELBO) of $\log p(\boldsymbol{x})$. Since $p(z)$ is assumed to be standard Gaussian distribution,

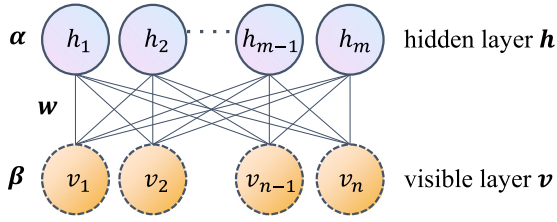

**Fig. 9.** The structure of restricted Boltzmann machine.

according to (60), $p(x)$ is only determined by $p(z|x)$. The model distribution $\tilde{p}(z|x)$ is an approximation of $p(z|x)$. When $\tilde{p}(z|x)$ approaches the real distribution $p(z|x)$, $\log p(x)$ is equal to the ELBO. Thus, it is reasonable to maximize ELBO instead of $\mathcal{L}$, since the latter cannot be calculated directly. ELBO can be computed as:

$$
\begin{aligned}
\text{ELBO} &= \int_z \tilde{p}(z|x) \log \frac{p(z,x)}{\tilde{p}(z|x)} \, dz \\
&= \int_z \tilde{p}(z|x) \log \frac{p(z)p(x|z)}{\tilde{p}(z|x)} \, dz \\
&= -KL(\tilde{p}(z|x) \parallel p(z)) + \mathbb{E}_{\tilde{p}(z|x)}[\log p(x|z)]
\end{aligned}
\tag{63}
$$

where the first term is related to KL divergence between $\tilde{p}(z|x)$ and $\mathcal{N}(\mathbf{0}, \mathbf{I})$, while the second term represents the expectation of $\log p(x|z)$ with the given encoder's output. Maximizing ELBO is equal to minimizing the KL divergence and maximizing the expectation. The second term can be considered as the negative reconstruction error of an ordinary AE (Vincent et al., 2010), and the KL divergence can be calculated as:

$$
KL(\tilde{p}(z|x) \parallel p(z)) = \frac{1}{2} \sum_{i=1}^{m} (-\log(\sigma_i^2) + \mu_i^2 + \sigma_i^2 - 1)
\tag{64}
$$

According to (64), this term is an extra KL loss that makes $\tilde{p}(z|x)$ as close as possible to the standard Gaussian distribution. Eliminating the KL loss, VAE will try to set all standard deviations to 0s such that the model can regenerate training data perfectly. In this situation, VAE is equal to an ordinary AE. In other words, the training of a VAE is an adversarial process between reconstruction loss and KL loss.

However, the sampling process is non-differentiable, so using the backpropagation algorithm directly is impossible. VAE uses a technique called the reparameterization trick to ensure $z$ is differentiable. To sample a latent vector $z$, it first samples $e = [e_1, e_2, \ldots, e_m] \in \mathbb{R}^m$ from a standard Gaussian distribution and calculates $z = \mu_z(x) + e \circ \sigma_z(x)$, which is differentiable.

#### 4.1.5. Other variants of autoencoder

Previous parts have introduced the classic AE and its popular variants. This part will present a brief discussion about other variants of AE.

In addition to the aforesaid popular AE variants, another well-known counterpart is the sparse AE (Ng et al., 2011). Sparse AE enables the hidden layer to have higher dimension. Only a few neurons are activated while most of them are limited to 0. This property comes from a penalty in the target function, which is often a KL divergence loss or L1 norm. A variant of sparse AE is k-sparse AE (Makhzani & Frey, 2013), which only keeps k highest activated neurons. With the development of CNN, convolutional layers are also adopted in AEs to replace fully connected layers. This architecture is called convolutional AE (Masci, Meier, Ciresan, & Schmidhuber, 2011). Like sparse AE, contractive AE (Rifai, Vincent, Muller, Glorot, & Bengio, 2011) also adds a penalty which is the Frobenius norm of inputs' Jacobian matrix. This regularization can suppress the turbulence among all the directions of input data. There are also some variants from the combination of different AEs. For example, denoising variational AE (Im Im, Ahn, Memisevic, & Bengio, 2017) combines DAE and VAE, which uses corrupted data as the input of VAE. The integration of DAE and sparse AE will yield the denoising sparse AE (Qiu, Zhou, Yu, & Du, 2018).

### 4.2. Industrial applications of mainstream DLVMs

DLVMs are widely used in many industrial areas. Among various application scenarios, we select two most typical fields — *process monitoring* and *soft sensor* — to present detailed discussions.

#### 4.2.1. Applications in process monitoring

As discussed, the original forms of AE and DBN are unsupervised models, which makes them quite suitable for unsupervised fault detection and diagnosis. AE and DBN are often used for unsupervised fault detection and diagnosis. Yu and Zhao (2019) proposed an exponential discriminant analysis (EDA) algorithm based on denoising AE (DAE) and elastic net (EN). First of all, EN algorithm is applied to DAE to a generate sparse network for process monitoring, and then the within-class and between-class covariance matrices of the values of the hidden neurons are calculated. Next, the EDA algorithm is used to get the optimal projection direction, and the hidden neurons corresponding to the nonzero coefficients are considered to be associated with the fault. Hallgrímsson, Niemann, and Lind (2020) proposed a strategy to generate sparse networks by minimization of the weight-redundancy optimization function and the pruning function. The SPE statistic of the reconstructed observation generated by sparse AE and the original observation is used to evaluate the process condition. If the SPE exceeds the computed control limit (threshold), the process is considered to be under abnormal situation. Then the contribution of each process variable to the SPE is calculated to analyze and isolate abnormal process variables. Besides, Hallgrímsson, Niemann, and Lind (2021) further set the contribution to a non-square form. The shift of reconstructed variables can be back-propagated through sparse AE to obtain the shift of original variables, so that a series of causal paths can be obtained. The causal paths that describe shifts of the different reconstructed variables with the same derived shifts of the original variables are combined into a reasoning path. Finally, the prior knowledge of process variables is used to screen out the correct reasoning paths: Cheng, Cai, Liao, Wu, and Dubey (2021) proposed a novel data driven fault isolation method based on the deep AE. The DAE model is learned by utilizing the unlabeled normal data. When the reconstruction error exceeds the threshold, it is considered that the data belongs to the abnormal class. Then the proportion of the reconstruction error is applied to find the abnormal source variables for fault isolation, and the variable with a greater proportion is considered to cause the fault. Liang, Duan, Bennett, and Mba (2020) develop a fault isolation scheme based on the sparse AE (SAE). Mahalanobis distance (MD) is used to calculate the distance between the monitoring data and the normal data. The faults can be detected by the changes of the reconstruction error. A two-dimensional contribution map is utilized for fault isolation and the variables with greater contributions generally indicate the fault source. Shao, Jiang, Li, and Liang (2018) proposed a continuous DBN (CDBN) for fault detection. First, a new comprehensive feature index is constructed using the locally linear embedding approach. Secondly, multiple trained continuous restricted Boltzmann machines are used to build the proposed CDBN. Finally, the essential parameters of the CDBN are optimized using the genetic algorithm (GA).

AE and DBN can also be modified to a supervised or semi-supervised form for fault detection and diagnosis. Li, Chai, and Yin (2021) proposed an AE-embedded dictionary learning method for fault diagnosis of nonlinear industrial process. The AE is used to map the original data to a nonlinear high-dimensional space, and then the dictionary is learned from the high-dimensional hidden features. In the process of learning parameters of nonlinear mapping and dictionary, two supervised graphs containing some prior knowledge of original data are introduced into the objective function. The framework of their proposed approach is shown in Fig. 16. Yang et al. (2021) proposed a multi-head deep neural network based on the sparse AE to realize fault detection and diagnosis. First, the encoder uses multiple layers of nonlinear transformations to extract high-level representations from

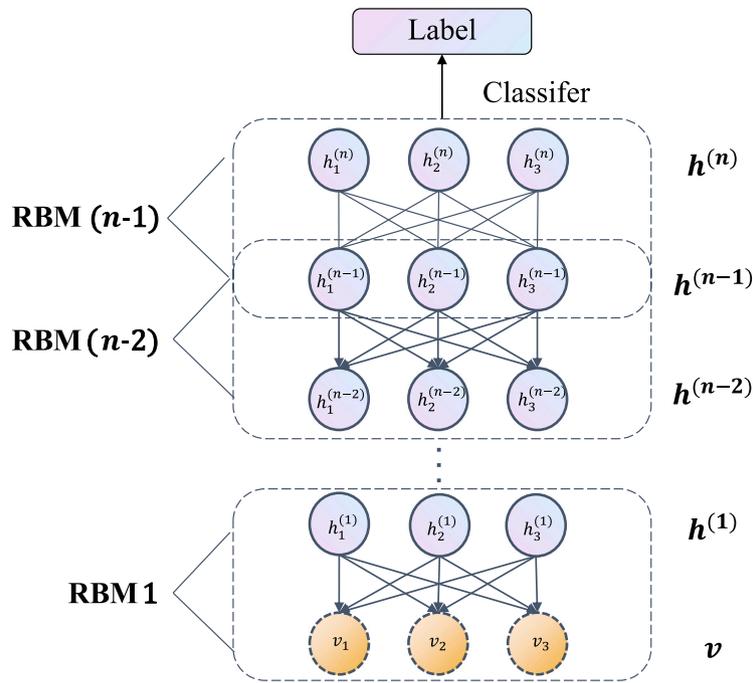

**Fig. 10.** The structure of deep belief network.

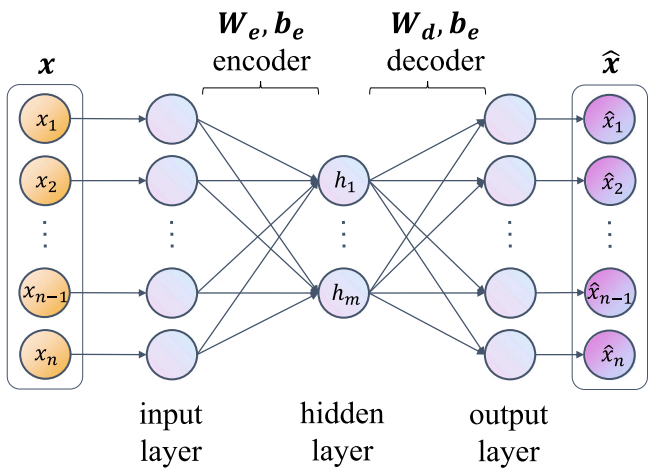

**Fig. 11.** The structure of classic AE.

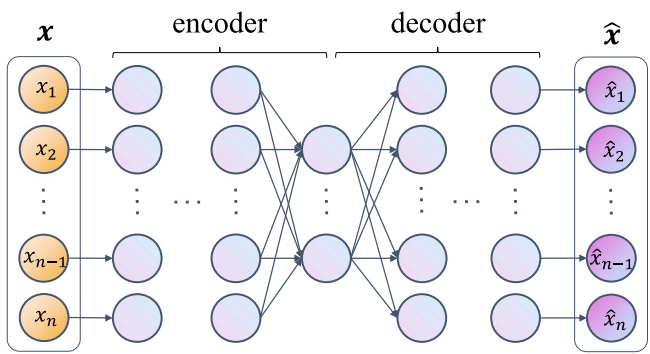

**Fig. 12.** The structure of deep AE.

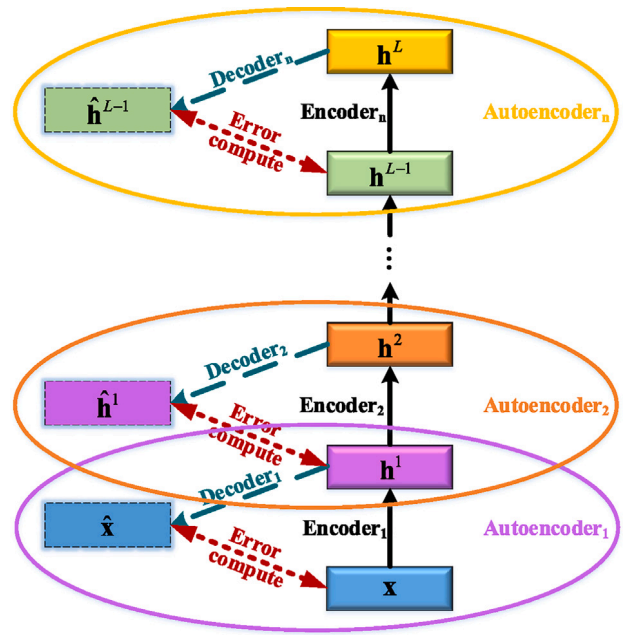

**Fig. 13. Model structure and learning process of SAE** (Sun & Ge, 2020). The leftmost vectors $(\hat{x}, \hat{h}^1, ....)$ are the added fake layers.

extracted features are shared as inputs to the decoder and the classification module. The decoder outputs the reconstructed observations and the classification module predicts the label of the input data. Kim et al. (2020) proposed a semi-supervised AE with an auxiliary task (SAAT) for fault diagnosis of the power transformer using dissolved gas analysis (DGA). The AE in the SAAT approach can both achieve unsupervised reconstruction and supervised identification through the shared hidden layers. In addition, by adding an auxiliary supervised fault detection task to the loss function of SSAE, the shared hidden layers can output the health features that can represent the health degradation properties. The architecture of their proposed network the input data, and the sparsity regularization is used to encourage the extraction of discriminative features from all hidden layers. Next, the

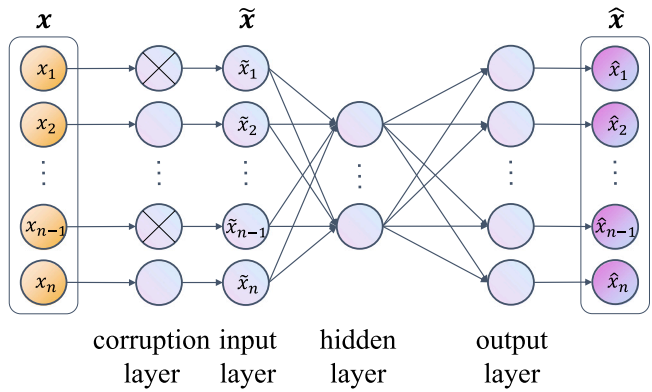

$x$      $\tilde{x}$      $\hat{x}$

corruption input    hidden    output
layer   layer    layer    layer

**Fig. 14.** The structure of denoising AE.

is shown in Fig. 17. Cui, Wang, Lin, and Zhong (2020) proposed a feature distance stacked AE (FD-SAE) for the fault diagnosis. SVM is applied for the fault detection and then FD-SAE is applied to classify the types of the faults. The information about the feature distance is added to the training process of the FD-SAE so that the features of the data are conducive to the classification. Pan, Wang, Yuan, Yang, and Gui (2021) proposed a classification-driven neuron-grouped stacked AE (CG-SAE) for fault detection. First, the hidden neurons of each AE are separated into many groups to introduce the information of the fault types. After that, each group of neurons mainly has a connection with a certain fault type in order to make the average activation of this group equal to the corresponding label of this type. Li, Li, Tong, and Zhang (2021) proposed a fault isolation method based on stacked AE (SAE) and the decision tree. The reconstruction model of the normal state is built using SAE to learn the historical state variables. The fault detection is performed based on the reconstructed errors of the state variables. Then, the relationship of the reconstruction errors in each state variable is extracted by the decision tree, and the fault isolation location is realized according to the relationship. Wang, Pan, Yuan, Yang, and Gui (2020) designed an extended DBN (EDBN) for fault classification. First, multiple extended restricted Boltzmann machines (ERBM) are used to extract the features in the pre-training phase. Then, a SoftMax layer is added to the final hidden layer of the last ERBM for the prediction of the fault types in the fine-tuning phase. Finally, the dynamic information is introduced to construct a dynamic modeling framework for fault classification. Deng, Liu, Xu, Zhao, and Song (2020) proposed an improved quantum-inspired differential evolution with multi-strategies, namely MSIQDE algorithm to construct an optimal DBN model for fault classification. Four strategies, the Mexh wavelet function, standard normal distribution, adaptive quantum state update and quantum nongate mutation combined with the original QDE algorithm are used to optimize the parameters of the DBN. Zhang and Zhao (2017) proposed an extensible deep DBN for fault classification. First, A variable selection algorithm is used to select variables according to the mutual information of the variables. Then, the sub-networks of the DBN is utilized to extract the features in both spatial and temporal domains. Finally, a global network of two layers is constructed for fault classification.

Robustness has always been a key issue in process monitoring, especially when the evaluated industrial process is quite complex. Learning the stable structure from uncertain data and exploring the relationship between process variables under normal operations can enhance the robustness of the model and effectively improve the performance of fault detection. Jiang, Song, Ge, and Chen (2017) added the robustness criterion to AE to carry out fault detection in industrial process. Firstly, the original normal training data is corrupted with Gaussian noise, and then the AE is trained so that the original uncorrupted normal data can be reconstructed approximately. Next, the square prediction error (SPE) in the training process is calculated and the control limit is determined. Finally, the fault detection is realized by calculating the SPE and comparing it with the control limit. Xiao, Lu, Wu, and Ai (2019) proposed a power quality (PQ) fault classification algorithm based on the stacked sparse denoising AEs (SSDAEs). First, the space vector transformation is used to convert the preprocessed PQ waveforms into 2D binary vectors. Then, the vectors are used as the inputs of the SSDAEs combined with supervised training and the parameter space is refined using the labeled data to improve the ability of discrimination. Yan, Guo, Li, et al. (2016) proposed a nonlinear robust process monitoring method based on contractive AE. Jacobian term is added to the objective function of the traditional AE to increase the robustness of the hidden representations to small perturbations around the original input. Specifically, contractive AE encourages the hidden representations to be contractive in the neighborhood of the original input by forcing the first derivatives of the hidden representations to be zero or small values. Peng, Zhang, Wang, and Zhang (2021) proposed a fault detection framework based on denoising sparse AE. Firstly, the noise is added to the original data, and then sparse parameters are predefined and the KL divergence is used to generate a sparse network, through which the original data is reconstructed. Finally, the hidden feature representations of the faults extracted from the denoising sparse AE finally output fault detection results through the SoftMax layer. Zhou, Ren, and Li (2020) proposed a robust copula double-subspace (CDS) model based on a sparse robust AE (SRAE) for fault detection. First, the training data is added with Gaussian noise and used as the input of the RAE. Second, the SAE is applied to extract high-dimension features. Then, the margin distribution subspace and the dependence structure subspace are obtained by dividing the extracted features. Finally, the joint margin distribution model and the copula model are established in each subspace, respectively. Zhu, Shi, et al. (2021) proposed a novel fault detection method based on the load weighted denoising AE (LWDAE). At first, the original loading matrix is obtained using the DAE. Then the loading matrix is weighted based on the magnitude and the direction of the data. In addition, two regularizations are added to prevent the impacts of noise and outlier. Prosvirin, Piltan, and Kim (2020) proposed a novel fault detection method based on a deep undercomplete denoising AE (DUDAE). First, the DUDAE is trained using the collected signals to approximate a nonlinear function of a system. Then, the residual signals are obtained using AE. Finally, the residual signals are used as inputs of the DNN for fault detection. Shi, Chen, Si, and Zheng (2020) proposed a new fault diagnosis algorithm based on a residual dilated pyramid network and full convolutional denoising autoencoder (RDPN-FCDAE). First, the data is used as the input of the encoding network during the unsupervised pre-training phase. Then the pre-trained encoding network and classification network are integrated into residual dilated pyramid full convolutional network (RDPFCN) during the fine-tuning phase.

The above methods deal with uncertain data from the perspective of robust modeling. On the other hand, the uncertainty of data can also be addressed from the stochastic and probabilistic perspectives. The variational AE (VAE), a probabilistic variant of the AEs, has a good probabilistic explanation for the generation of the reconstructed observations from a specific distribution. The deep network structure of VAE can approximate complex posterior and conditional distributions to infer latent variables and generate reconstructed observations. Wang, Forbes, Gopaluni, Chen, and Song (2019) proposed a fault detection algorithm based on variational AE for complex nonlinear processes. First, the observed variables are mapped to the latent variables by the encoder to calculate the posterior distribution of the latent variables, and then the samples are drawn from the calculated distribution and transferred to the decoder to calculate the conditional distribution of the observed variables which are reconstructed by the decoder. Wang, Yuan, Chen, and Wang (2021) proposed a fault detection scheme based on aprobabilistic generative deep learning model developed from supervised and semi-supervised variational AEs (SVAE and $S^2$VAE). In SVAE, the distributions of the reconstructed observations of the process variables and the quality variables are the Gaussian distributions. $S^2$VAE added a predictor network on the basis of SVAE, and the labeled data is used for supervised training to obtain the conditional distribution of the quality variables under process variables. Lee and Chen (2021) proposed a multi-grade processes fault detection framework based on Gaussian mixture prior variational AE. The samples in the target and the source datasets are generated from a grade variable and some latent variables. The data from the source and the target grade and the data only from target grade are the inputs of the encoder. And the output of the encoder is a GMM posterior distribution consisting of Gaussian posterior distributions of latent variables from different source grades data and the target data. Zhang, Su, Meng, and Dong (2020) developed a fault detection approach based on the denoising variational AE (DVAE). The observations added with the noise are used to train the proposed DVAE. The latent space mapping based on the noise-corrupted data can increase the robustness of the inference model. The DVAE generates the outputs in the form of a distribution, which not only predicts the values of the latent variable but also deals with the uncertainties. Chen, Mao, Zhao, Jiang, and Zhang (2020) proposed a variational stacked AE (VSAE) for fault detection. The VAE is added into the SAE to construct the VSAE. First, two AEs are used to extract the features of input signals. Then, a VAE network is utilized to model the distribution of latent feature space.

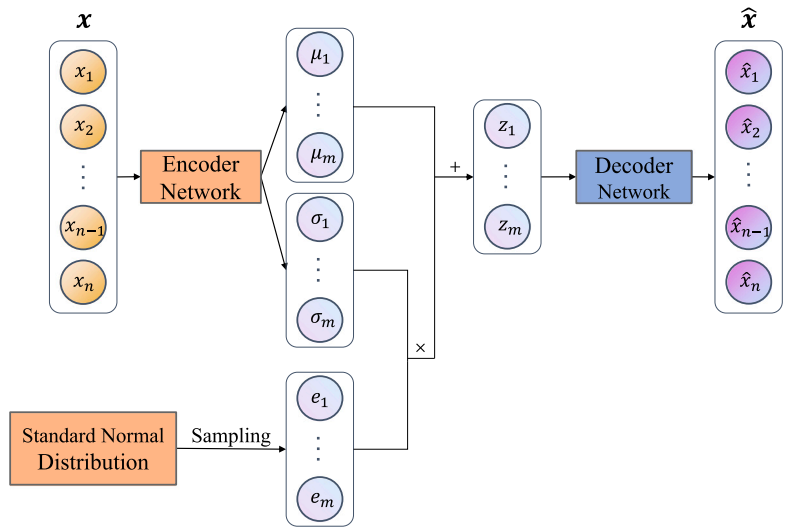

**Fig. 15.** The structure of variational AE.

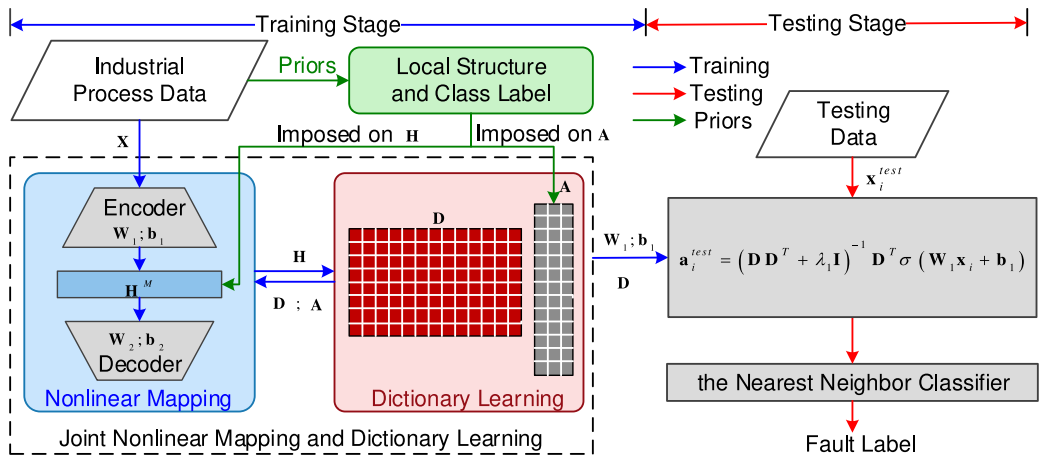

**Fig. 16. The AE embedded dictionary learning framework** (Li, Chai, & Yin, 2021). AE is adopted to map the original process data to a high-dimensional feature space. Dictionary learning is used to extract inherent structure in data. Then the nearest neighbor classifier is utilized to predict the fault label.

The harmony search optimizer algorithm is applied in order to adjust the structure of the network adaptively according to various datasets. He and Jin (2021) developed a novel fault detection model called deep variational AE classifier (DVAEC). First, the latent variables containing uncertainties are learned using the variational AE (VAE) network so that the hidden features of the input data can be extracted. Then, the learned latent variables combined with their distributions are used as the inputs of a DNN for fault classification. The architecture of their proposed framework is shown in Fig. 18. Chen, Liu, Xia, Wang, and Lai (2020) proposed a sliding-window convolutional variational autoencoder (SWCVAE) for fault detection. The encoder and the decoder are both composed of three convolutional layers and a flatten layer. The final hidden states outputted from the encoder and the decoder are used to obtain the posterior distribution and the decoding distribution using linear layers and softplus layers, respectively.

There are also complex dynamic processes in the industrial scenarios, and the dynamic behavior exists among process variables. Therefore, the dynamicity is another important issue that has a significant influence on the fault detection. The temporal information exists in the time-series data, and the state of the current industrial variables is affected by the previous and current events. Zhang, Zhu, Ge, et al. (2020) proposed an end-to-end trainable framework based on the recurrent Kalman variational AE for process modeling and fault detection. Firstly, the variational AE is used to encode the time series data to obtain the latent Gaussian representations. Then, a series of locally linear models are constructed to linearly approximate the nonlinear dynamic parameters and LSTM is used to update the recurrent parameters. Next, the Kalman filter is used to integrate the dynamic inference process to estimate the hidden state uncertainty due to process noise. Yu and Yan (2020) designed a multiscale fault detection system based on stacked denoising AE (SDAE). Firstly, the multiscale analysis is carried out on the basis of agglomerative hierarchical clustering and silhouette coefficient to extract the feature information of the original data in different scales. Then SDAE models are built at each scale to learn high-order and robust features from the noise data, and Bayesian inference technique is used to integrate the calculated results of monitoring statistics at different scales. Yin and Yan (2019) proposed a nonlinear dynamic fault detection model based on mutual information (MI) and stacked sparse AE (SSAE). First, MI is used to select historical data of the past moment that is highly correlated with the data at the current moment. Next, the current moment data and the most relevant past moment data are combined as the inputs of SSAE to extract dynamic and nonlinear features. Bi and Zhao (2021) proposed orthogonal self-attentive variational AE for fault detection and identification. A spatial self-attention layer and a temporal self-attention layer are used to construct the orthogonal attention model. The self-attention mechanism is introduced into the VSAE to integrate information from all times and the output of the orthogonal attention model is reconstructed through the VSAE. Han, Ellefsen, Li, Holmeset, and Zhang (2021) proposed a LSTM-based variational AE (LSTM-VAE) for fault detection. The values measured by multiple sensors are mapped into a latent space through the encoder at each time step, and the values are reconstructed from the latent space through the decoder. The LSTM is introduced into the encoder and decoder to aggregate the information across each time step. The anomaly score is calculated using the log-likelihood of the current reconstruction given the expected distribution. Yang and Zhang (2020) proposed a joint variational AE (JVAE) for fault detection. First, time series with multiple variables are selected from the dataset and are used as inputs of the proposed network. Next, four essential parameters are chosen as the input parameters and they are separated into two categories. Two categories of the parameters are reconstructed using two variational AEs, respectively.

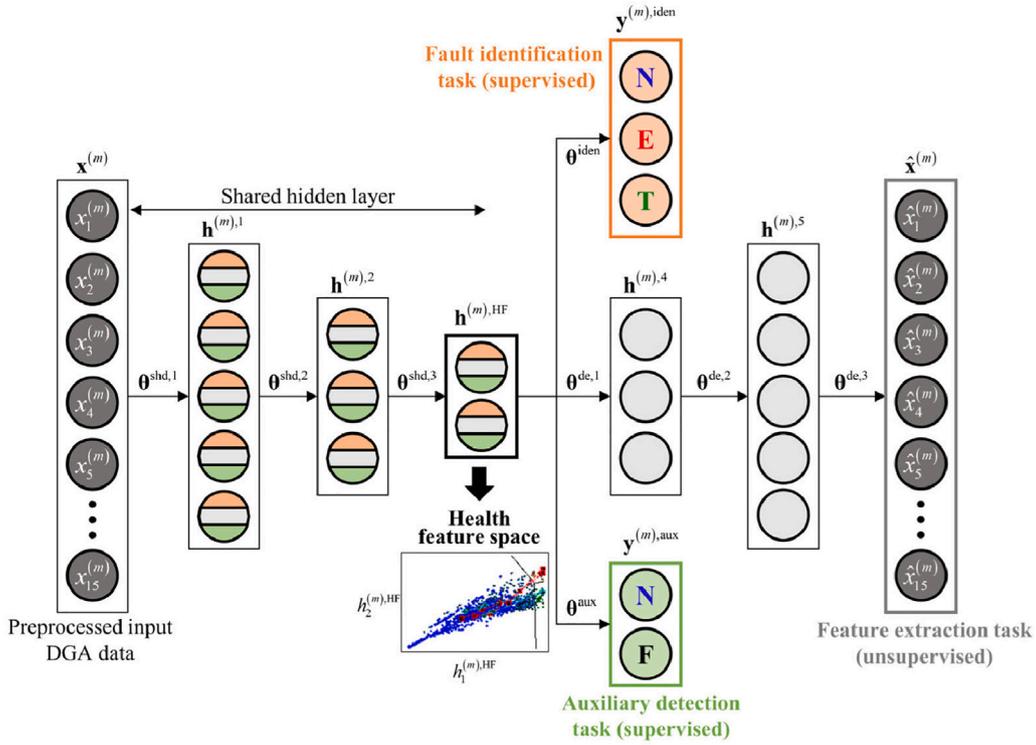

**Fig. 17.** Model architecture of SAAT (Kim et al., 2020). Orange, gray, and green shared hidden layers stand for the features related to the fault identification, representative characteristics of DGA data, and health degradation.

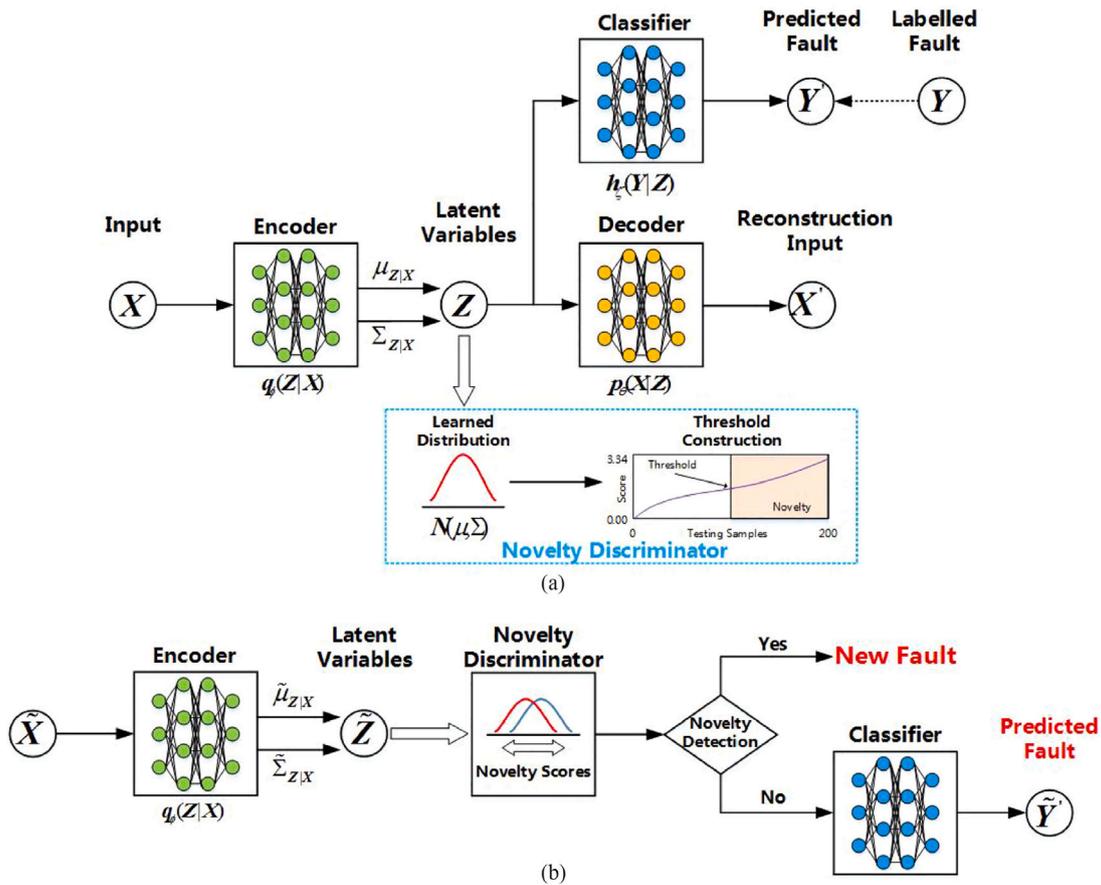

**Fig. 18.** Model architecture of DVAEC (He & Jin, 2021). In the offline training phase, DVAEC is designed to train the fault classifier and VAE. In the online testing stage, DVAEC performs fault identification.

### 4.2.2. Applications in soft sensor

Since soft sensor modeling is essentially a regression problem, AE and DBN in this area are often modified to a supervised or semi-supervised form to complete the predictive tasks. Yuan, Huang, Wang, Yang, and Gui (2018) proposed a variable-wise weighted stacked AE (VW-SAE) to predict the key variables. The proposed model calculates the correlation coefficients between input variables and output variables and assigns different weights to the input variables. Then, the pre-trained autoencoders are stacked to form a deep network, and the parameters are fine-tuned in the whole network. Lian, Liu, Wang, and Guo (2019) proposed a soft sensor method based on DBN. First, the grey relation analysis method is applied to select the variables. Then, the labeled data and the unlabeled data are used to train the DBN which can extract features from the dataset. Finally, the features are used as the inputs of the support vector regression (SVR) for prediction. Yuan, Gu, Wang, Yang, and Gui (2019) proposed a soft sensor modeling framework based on the stacked supervised AE. A network of three layers is built during the pre-training phase of each hidden layer, in which the encoder extracts the high-order features and the decoder outputs the prediction of the quality value. A SoftMax layer is added to the final layer of the whole network to predict the quality value. Yuan, Gu, and Wang (2020) proposed a novel supervised DBN (SDBN) for soft sensor modeling. First, the target variables are added into the visible layer of RBM to build supervised RBM (SRBM). Then, the multiple SRBMs are stacked to establish SDBN to extract the deep features that are related to the target variables. Tanny, Chen, and Wang (2020) proposed a novel differential entropy soft sensor model based on the variational AE (VAE). First, the process variables are used as the inputs of the prediction network to predict target variables and the entropy of the prediction are used as the inputs of the VAE. Yuan, Ou, Wang, Yang, and Gui (2019) develop a layer-wise data augmentation (LWDA)-based stacked AE for soft sensor modeling. New synthesized samples are obtained using the linear interpolation technique and they are combined with the raw samples during the pre-training phase. Then, the target variable is fed into the final layer of the network for the fine-tuning of the parameters. Yuan, Zhou, et al. (2019) proposed a stacked quality-driven AE (SQAE) for soft sensor modeling in industrial processes. The target variables are added to the output layer of each QAE, so the extracted hidden features are related to the target variables. Then, a regression layer is established on the final layer of SQAE to achieve the prediction of the target output. The architecture of their proposed network is shown in Fig. 19. de Lima and de Araujo (2021) proposed a novel ensemble deep relevant learning soft sensor (EDRLSS) modeling framework based on stacked AE (SAE). The mutual information (MI) between quality variables and extracted latent features is introduced to train the proposed network hierarchically. Besides, a bagging-based approach is utilized in the supervised fine-tuning phase. Liu, Wang, Ye, Wang, and Yuan (2021) proposed a stacked neighborhood preserving AE (S-NPAE) for soft sensor modeling. The similarity matrices are computed according to the adjacent geometrical graph of the input variables established by KNN algorithm. Then, the proposed network is restricted using the regularization. After the layer-by-layer pre-training phase, a regression layer is added to the final layer of the network during the fine-tuning phase. Liu, Wang, et al. (2021) proposed a nonlocal and local structure preserving stacked AE (NLSP-SAE). The nonadjacent and adjacent weight matrices are computed according to the adjacent graph established on the input variables. Then, they are added to the objective function with the global Euclidean variance of the data for modeling.

AE-based and DBN-based local learning models have also been widely applied for soft sensor modeling. Guo, Bai, and Huang (2020) proposed an output-relevant variational AE for soft sensor modeling in industrial processes. Firstly, the correlation coefficients between each dimension of input variables and the target variable are calculated, and different weights are assigned to the input variables. Next, the historical samples that are most relevant to the query data are selected for Gaussian process regression (GPR) local modeling. Guo and Huang (2020) further proposed an output-relevant variational AE based on the mutual information (MI). The input variables that have nonlinear relationships with the quality variables are chosen according to MI. The Symmetric Kullback–Leibler (SKL) divergence is applied to select the historical input data that are related to the query data for local modeling. The outliers and the missing data are coped with using the expectation maximization (EM) algorithm. Guo, Wei, and Huang (2021) developed a Gaussian mixture VAE (GMVAE) to construct a just-in-time learning (JITL) framework. The features extracted through GMVAE follow a Gaussian mixture distribution. The historical samples are chosen by computing the mixture symmetric Kullback–Leibler (MSKL) divergence and then a mixture probabilistic principal component regression (MPPCR) model is utilized for local modeling. Lyu, Chen, and Song (2021) proposed a regressor-embedded semi-supervised variational AE (RSSVAE) for soft sensor modeling. The unlabeled data in the data-scarce region and the labeled data which has similar information with that in the data-scarce region are both used to obtain the synthetic labeled data. Then, they are combined with the original data for local soft sensor modeling. Wu, Liu, Yuan, and Wang (2020) proposed a just-in-time framework based on the stacked AE (JIT-SAE) for quality variable prediction. First, the proposed model is trained with the historical data during the pre-training and the fine-tuning phase. Then, the historical samples that are most relevant to the query data are chosen and they are weighted based on the similarity with the query data during the online fine-tuning phase. Liu, Shao, and Chen (2020) proposed an AE-based nonlinear Bayesian weighted regression (NBWR) for soft sensor modeling. First, the hidden features of the historical samples are extracted through the AE. Then, the similarities between the features of the query sample and the historical sample are computed. The most relevant historical samples are chosen and different weights are assigned to them to build a local Bayesian regression model. Qiu, Wang, Zhou, Guo, and Wang (2020) proposed a novel soft sensor framework called semisupervised just-in-time relevance vector regression (SJRVR). First, the unlabeled data is processed based on the data generation capability of adversarial AE. Then, the process variables are separated into multiple phases through the Gaussian mixture model. The historical samples are chosen according to a similarity measurement based on spatial distance and phase distance. Finally, the local relevance vector regression model is trained for prediction. Yuan et al. (2021) proposed an online adaptive fine-tuning of DBN (OAFDBN) for quality prediction. First, the DBN model is trained using the historical dataset during the offline pre-training and fine-tuning phase. During the online prediction, the historical samples that are most relevant to the query data are chosen to fine-tune the model online.

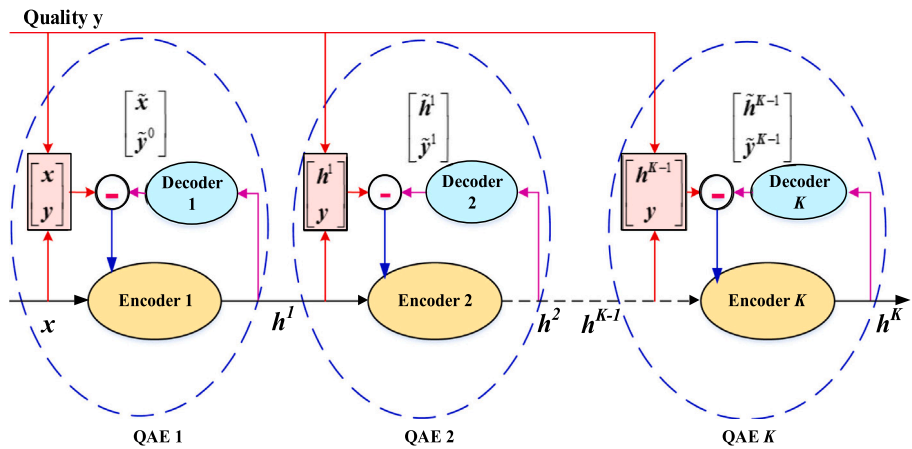

**Fig. 19. Model architecture of SQAE** (Yuan, Zhou, et al., 2019). The raw input data and the quality data are simultaneously reconstructed at the output layer of the QAE. The feature vectors of the previous layer are connected to the input layer of the next QAE.

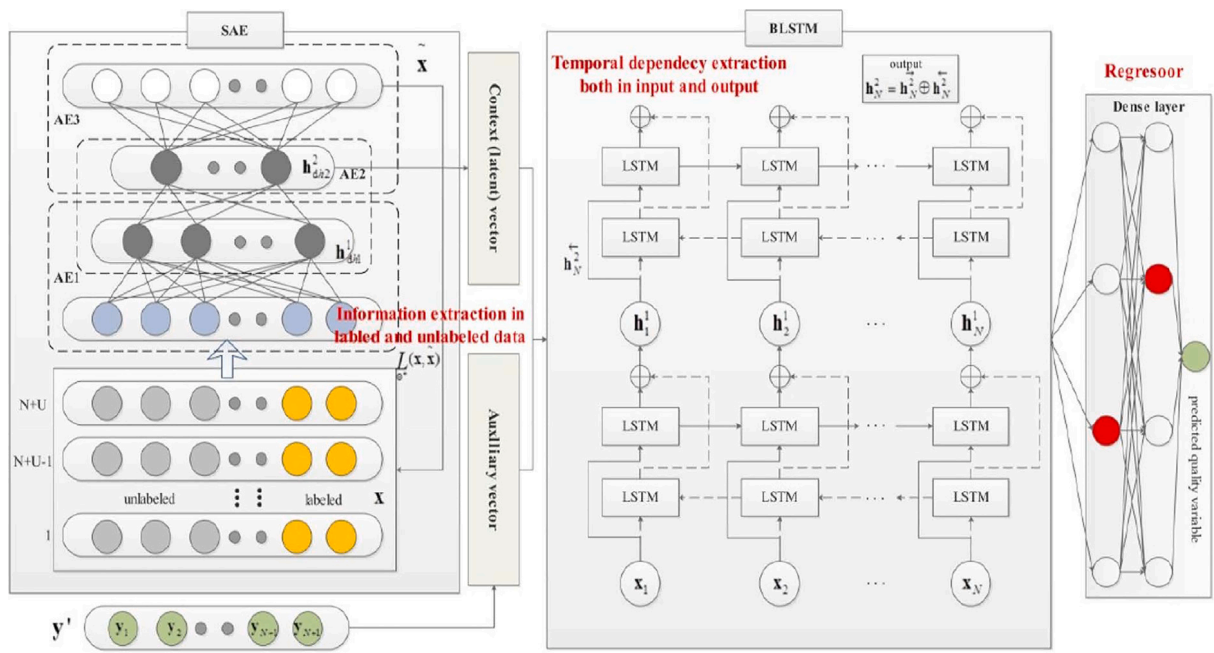

**Fig. 20. Model architecture proposed in** Yin, Niu, He, Li, and Lee (2020). SAE is capable to learn high-level abstract features from labeled and unlabeled data. Then, the time dependencies of labeled context vector can be captured by deep BLSTM network. A regression layer is applied for quality prediction.

The dynamicity exists in many industrial processes and AE-based models with dynamic information have been derived for dynamic processes. Lee, Ooi, Tanny, and Chen (2021) proposed a latent dynamic variational AE (LDVAE) for soft sensor modeling. First, the input data sequences are mapped into the latent space by the encoder. Then, the encoded variables are used as the inputs of the bi-directional RNN to generate the filtered distributions and the smoothed distributions. Finally, the input data sequences are reconstructed through the decoder. Moreira de Lima and Ugulino de Araújo (2021) proposed a novel dynamic soft-sensor modeling approach based on the stacked AE (SAE) and LSTM. The mutual information (MI) is applied to compute a correlation coefficient during the unsupervised pre-training phase of the SAE. The LSTM is added to the final layer of the SAE to model the dynamic behavior during the supervised fine-tuning phase. Xu and Cai (2021) proposed a novel Gaussian Mixture GRU-based variation AE regression (GM-GVAER) model. Gated Recurrent Unit (GRU) cells are used to extract dynamic features from the data sequences. The Gaussian Mixture priors are applied to characterize the multiple phases of data. Yao and Ge (2021) proposed a dynamic features incorporated locally weighted AE regression (DFI-LWAER) network for soft sensor modeling. The LSTM encoder–decoder with attention mechanism extracts the hidden dynamic features and they are combined with the original inputs to take part in the similarity measurement and local modeling. Shen and Ge (2020b) proposed a weighted nonlinear dynamic system based on variational AE for soft sensor modeling. The maximal information coefficient (MIC) is used for extracting the output-related features by computing the importance of the variables from the variable correlation and the dynamic correlation. Shen and Ge (2020a) proposed a supervised nonlinear dynamic system based on variational AE for quality prediction. The latent variables at current moment are jointly determined by the original input variables at current moment and the latent variables at previous moment. Each latent variable is reconstructed through the decoder, and the output is obtained through the regression model. The constraint of the distributions of latent variables at previous moment is added to the original variational AE loss function. Xie et al. (2020) proposed a variational AE bidirectional LSTM soft-sensor model based on batch training (Bt-VAEBiLSTM). First, the training data is separated into multiple batches based on time series. Then, the data is reconstructed through the variational AE according to their probability distribution. Finally, they are used as the inputs of the bidirectional LSTM for time series modeling. Yin et al. (2020) proposed a soft sensor modeling method based on the stacked AE (SAE) and the bidirectional LSTM (BLSTM). The hidden features of the input data are extracted through the unsupervised SAE and they are converted to the context vector. Then, they and quality value at previous moment are incorporated into a new vector. Next, the vector is used as the input of BLSTM to extract the features. Finally, the features are used as the inputs of a regression layer for quality prediction. The architecture of their proposed model is shown in Fig. 20. Chen and Ge (2021) proposed a novel framework incorporating LSTM, graph convolution and SAE to predict quality variables. They first adopt the spatial self-attention mechanism and LSTM to discover knowledge from process variables, which is essentially a graph. Then graph convolution is utilized to further process the extracted knowledge. Finally, the features output from the graph convolution layers are fed into SAE to make predictions. The architecture of their framework is shown in Fig. 21.

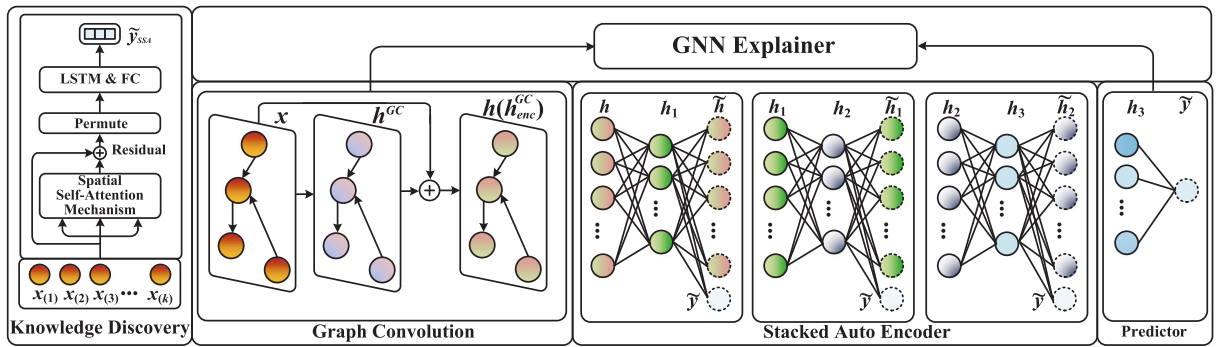

**Fig. 21.** Model architecture proposed in Chen and Ge (2021). The attention mechanism and LSTM are applied to discover knowledge, which is further processed through graph convolution. The features output from the graph convolution layers are used as inputs of SAE for quality prediction.

### 4.2.3. General application status

In general, the overall application status of five most commonly used DLVMs in the whole industrial modeling field is shown in Fig. 22(a), where SpAE represents sparse AE. As can be seen, the most popular DLVMs in the industry community are DBN and SAE. Fig. 22(b) displays the publishing trends of the aforesaid five popular DLVMs in the industrial modeling area. It is obvious that publications based on DLVMs have a significant growth since 2015, especially after 2018, which is very consistent with the publishing trends of deep learning.

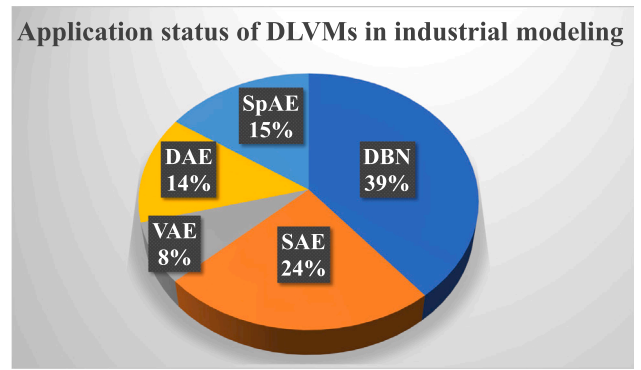

(a) General application status of DLVMs in industrial modeling.

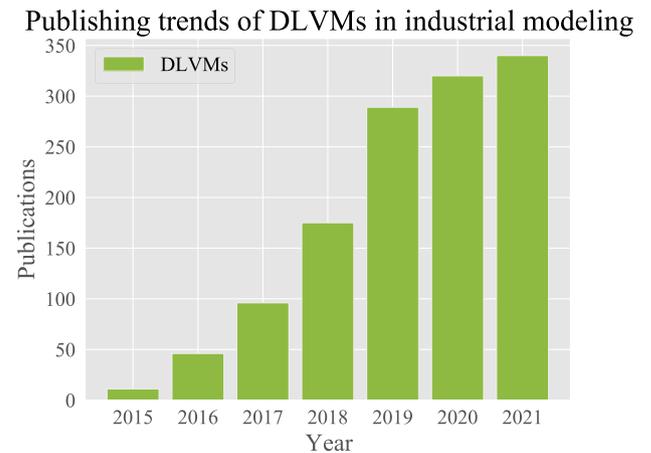

(b) Publishing trends of DLVMs in industrial modeling (2015-2021).

**Fig. 22.** Some application statistics of five most commonly used DLVMs in industrial modeling.

## 5. Lightweight deep LVMs

This section first discusses the motivation and connotation of LDLVM, then provides two novel LDLVMs, along with detailed descriptions of their principles, architectures and merits.

### 5.1. Motivation, definition and connotation of LDLVM

We have presented detailed discussions on traditional LVMs and deep LVMs in Sections 3 and 4, respectively. As mentioned in Section 1,

the advantages and disadvantages of the two kinds of LVMs are both obvious. Traditional LVMs have better model interpretability, fewer parameters, shorter running time, and they are not severely dependent on the data volume, but their performance cannot cope with complex industrial process modeling. Current neural networks-based DLVMs can achieve better results on many modern industrial modeling tasks, but it also comes with sacrifices in model interpretability and modeling efficiency, meanwhile they rely too much on the data volume, which means their performance may plummet when the sample size is relatively small.

Is there a way to neutralize the characteristics of these two kinds of LVMs? Or in other words, is there a way to combine their virtues, i.e., building a *high-performance* LVM meanwhile maintaining *good interpretability* and *high modeling efficiency*? Or more generally, current DLVMs are all neural networks-based deep models, are there other ways to build a deep model? As can be seen from Section 4, DLVMs use the hierarchical model structure to increase model complexity so that to improve the ability of representation and feature extraction, which is crucial for improving performance. The development of ML has shown that model complexity is necessary for model performance, and the unique hierarchical model structure of DL is a good way to increase complexity. Based on the above ideas and analysis, we propose a new concept called lightweight deep LVM. LDLVM holds the basic idea — deep, which means it utilizes hierarchical and cascaded structures to increase model complexity so that to improve the ability of representation and feature extraction. But the fundamental modeling tools are not artificial neurons. Instead, LDLVM makes use of simpler approaches, such as the concise traditional LVMs, to build deep models. In other words, LDLVM is like deepening the traditional shallow LVMs. Since the traditional LVMs are usually transparent with good interpretability and efficient modeling procedures, by effectively deepening them to deep forms, we believe the LDLVM can enhance the performance of the corresponding traditional model (even achieve competitive performance compared with current mainstream deep models), while keeping the original merits of traditional LVM that current deep models do not have.

Generally, based on the aforesaid discussions, the motivations of LDLVM can be summarized as *(1) combining the virtues of traditional LVMs and DLVMs* and *(2) exploring non-neural-network manners to build deep models*. The connotation and purpose of LDLVM are to achieve competitive performance compared with current neural networks-based DL models in appropriate application scenarios, while overcoming the shortcomings of current deep models to the maximum. Such a connotation is vividly shown in Fig. 23 through a comic.

This subsection has discussed the motivation and connotation of LDLVM. In addition, we have also successfully built two novel LDLVMs which accord with the proposed connotation. These two LDLVMs target the *unsupervised fault detection* domain and the *supervised classification and regression* domain, respectively. The principles and architectures of these two models are discussed in Sections 5.2 and 5.3, respectively.

### 5.2. Unsupervised fault detection based on LDLVM

This subsection generally refers to Kong and Ge (2022b), which is our first work on exploring the idea of LDLVM. In short, this work has built an LVM-based deep fault detection framework through *hierarchical feature extraction*, *Bayesian inference* and *proper weighting strategy*. The framework can be applied to most traditional shallow LVMs for deepening them to deep forms so that to achieve better fault detection performance.

---

⁵ Known as LDM in the picture, for the relationship between LDLVM and LDM, please see footnote 1.

#### 5.2.1. Principles of the proposed framework

Specifically, the proposed framework has four main procedures.

*(a) Building hierarchical features and monitoring statistics:*

Let $f(\cdot)$ represents a traditional LVM that can be used for fault detection (such as PCA, KPCA or ICA). $\boldsymbol{x}(i) \in \mathbb{R}^m$ represents a sample vector with $m$ dimensions. The following steps are used to extract multiple layers of latent variables and transform them into monitoring statistics.

$$\boldsymbol{z}^1(i) = f(\boldsymbol{x}(i))$$
$$\boldsymbol{z}^l(i) = f(\boldsymbol{z}^{l-1}(i)), \; 2 \leqslant l \leqslant L$$
$$S^l(i) = g(\boldsymbol{z}^l(i)), 1 \leqslant i \leqslant n, \; 1 \leqslant l \leqslant L \tag{65}$$

where $n$ is the sample size, $L$ is the total layers, i.e. the model depth (see Figs. 24 and 25 for more visual representation of model depth). $g(\cdot)$ is a function to formulate monitoring statistic $S$. Through (65), at each layer, $n$ training samples can be turned into $n$ monitoring statistics.

*(b) Transforming monitoring statistics into posterior probabilities:*

Given $\boldsymbol{x}_t \in \mathbb{R}^m$ as a test sample. At the $l$th layer, denoting the conditional probabilities of sample $\boldsymbol{x}_t$ under fault state $F$ and normal state $N$ as $P_S^l(\boldsymbol{x}_t | F)$ and $P_S^l(\boldsymbol{x}_t | N)$, respectively. The following steps utilize Bayesian inference to transform the monitoring statistics into posterior probabilities.

$$P_S^l(\boldsymbol{x}_t | F) = \exp(-\mu S_{\lim}^l / S^l(t)), 1 \leqslant l \leqslant L$$
$$P_S^l(\boldsymbol{x}_t | N) = \exp(-\mu S^l(t) / S_{\lim}^l), \; 1 \leqslant l \leqslant L$$
$$P_S^l(\boldsymbol{x}_t) = P_S^l(\boldsymbol{x}_t | N) P_S^l(N) + P_S^l(\boldsymbol{x}_t | F) P_S^l(F), 1 \leqslant l \leqslant L$$
$$P_S^l(F | \boldsymbol{x}_t) = \frac{P_S^l(\boldsymbol{x}_t | F) P_S^l(F)}{P_S^l(\boldsymbol{x}_t)}, \; 1 \leqslant l \leqslant L \tag{66}$$

where $\mu$ is an adjustment parameter to decrease the impact of data outliers, $S_{\lim}^l$ is the control limit at $l$th layer, which is obtained through performing KDE on the monitoring statistics of normal samples. $P_S^l(F)$ (or $P_S^l(N)$) is the prior probability that the system is in fault (or normal) state, which equals to $\delta$ (or 1-$\delta$). $P_S^l(F | \boldsymbol{x}_t)$ is the obtained posterior probability that $\boldsymbol{x}_t$ is a fault sample.

*(c) Weighting the posterior probabilities at different layers:*

The $L$ posterior probabilities $P_S^l(F | \boldsymbol{x}_t)(1 \leqslant l \leqslant L)$ contain the fault information at different layers. A simple and effective way to combine the information is to weight the posterior probabilities based on suitable rules:

$$\bar{P}_S^l(F | \boldsymbol{x}_t) = \frac{\sum_{j=t-Y+1}^t P_S^l(F | \boldsymbol{x}_j)}{Y}$$
$$w_S^l = \begin{cases} \dfrac{1}{\eta}, \; \text{if } P_S^l(F | \boldsymbol{x}_t) \geqslant \delta \& \bar{P}_S^l(F | \boldsymbol{x}_t) \geqslant \delta \\ \eta, \; \text{otherwise} \end{cases}$$
$$\text{DBS} = \frac{\sum_{l=1}^L w_S^l P_S^l(F | \boldsymbol{x}_t)}{\sum_{l=1}^L w_S^l} \tag{67}$$

where $\bar{P}_S^l(F | \boldsymbol{x}_t)$ is a mean posterior fault probability based on the previous $Y$ samples of $\boldsymbol{x}_t$. $w_S^l$ is the proposed weighting strategy, and $\eta$ is a small positive number and $\delta$ is the preset significance level. DBS is the abbreviation of *deep Bayesian statistic*, which is a monitoring index that integrates all $L$ layers information extracted by the hierarchical structure. If DBS $\geqslant \delta$, the system is considered in a fault state. DBS < $\delta$ means there is no fault in the process.

*(d) Further integrating the outputs from different models:*

The framework can be applied to most traditional LVMs, thus will provide us several DBSs that come from different models. Actually, the above fusion strategy can be further utilized to integrate the DBSs from different models. Define the whole framework (i.e., (65)–(67)) as $R(\cdot)$. Then the generation of a more comprehensive probabilistic index named as overall deep Bayesian statistic (ODBS) can be expressed as

$$R(S_1^1, S_1^2, \cdots, S_1^{L_1}) = \text{DBS}_1$$

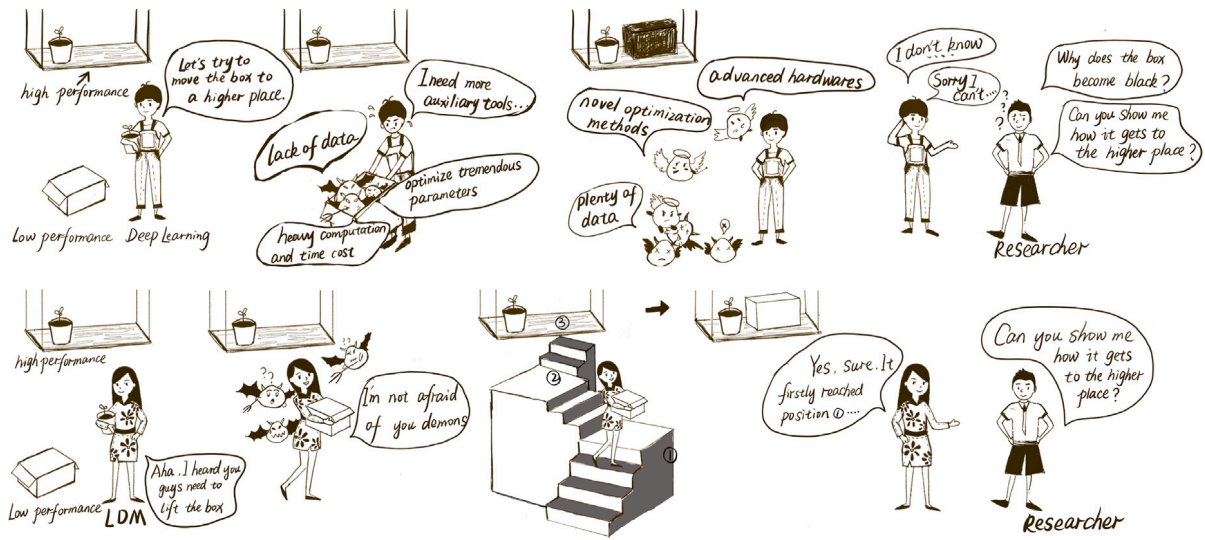

**Fig. 23. The connotation of LDLVM** (Kong & Ge, 2022c). In the upper part of the picture, current deep learning method tried to lift the box from "low performance" to "high performance". However, in this process, "deep learning" encountered various resistances, such as lack of data, heavy computation and so on. In order to deal with these resistances, "deep learning" needs some outside help, for example, advanced hardware, etc. After solving these problems through external help, "deep learning" successfully lifted the box up, but it had turned the original white box into a black box, and simultaneously could not explain how the box got up. On the other hand, the lower part of the picture tells us that LDLVM[5] not only has no problems such as lack of data, but also can clearly show researcher how it improves performance.

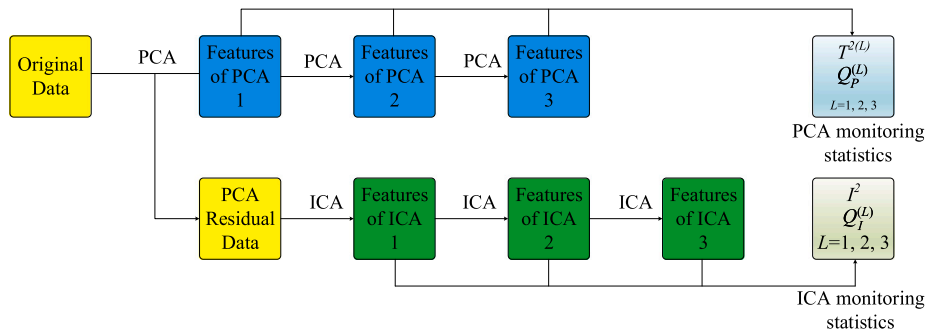

**Fig. 24. Offline training schematic of DPI-3L.** In the offline training stage, the dataset is projected into the model structure to calculate the mapping matrix and the projected features. The features of each model at each layer are used to compute monitoring statistics, and KDE is used to determine their control limits. The purpose of offline training is to obtain the mapping matrix and the control limits.

$$R(S_2^1, S_2^2, \cdots, S_2^{L2}) = \text{DBS}_2$$
$$\cdots$$
$$R(S_K^1, S_K^2, \cdots, S_K^{L_K}) = \text{DBS}_K$$
$$R(\text{DBS}_1, \text{DBS}_2, \cdots, \text{DBS}_K) = \text{ODBS} \quad (68)$$

where $K$ is the number of DBSs involved in generating ODBS.

For more detailed information about the principles of our proposed framework, please refer to Kong and Ge (2022b).

### 5.2.2. An instance of the proposed framework

To demonstrate the effectiveness of the proposed framework, we adopt two classical LVMs, PCA and ICA, to develop a deep model for fault detection. The principles of PCA and ICA are presented in Sections 3.2.2 and 3.2.3.

To provide an example, we develop a three layers deep PCA-ICA model (DPI-3L) based on the processes presented in Section 5.2.1, as illustrated in Fig. 24. This figure displays the offline training procedures of DPI-3L, through which we can obtain the projection matrices of different layers and the control limits of various indices. According to Fig. 24, the features of PCA at each layer are computed by

$$T^{(1)} = X P^{(1)}$$
$$T^{(l)} = T^{(l-1)} P^{(l)}, \ l = 2, 3, \ldots \quad (69)$$

Similarly, the features of ICA at each layer are calculated by

$$S^{(1)} = W^{(1)} E$$
$$S^{(l)} = W^{(l)} S^{(l-1)}, \ l = 2, 3, \ldots \quad (70)$$

where $E$ is the PCA residual matrix in the first layer.

Fig. 25 depicts the process of calculating monitoring statistics at different layers for a new test sample. After obtaining the statistics, fusion strategy (66)–(67) is used to transform them into probabilistic DBSs. DPI-3L generates four traditional monitoring statistics — $T^2$ and $Q_T$ in PCA, $I^2$ and $Q_I$ in ICA. Thus, the aforesaid process will produce four probabilistic statistics — DBT², DBQ$_P$, DBI² and DBQ$_I$, respectively. Finally, (68) is used to transform them to the ODBS index:

$$\text{ODBS} = R(\text{DBT}^2, \text{DBQ}_P, \text{DBI}^2, \text{DBQ}_I) \quad (71)$$

Through the above procedures, we have successfully built a LDLVM-based detection model and constructed a monitoring index DOBS. The superiority of the proposed model are presented in Kong and Ge (2022b) though evaluations on the TE process (Downs & Vogel, 1993) (a benchmark for fault detection). Results show that although the proposed model involves only linear transformations, its performance has surpassed many nonlinear methods, even the DNN-based deep models. Besides, the proposed model also has several other advantages,

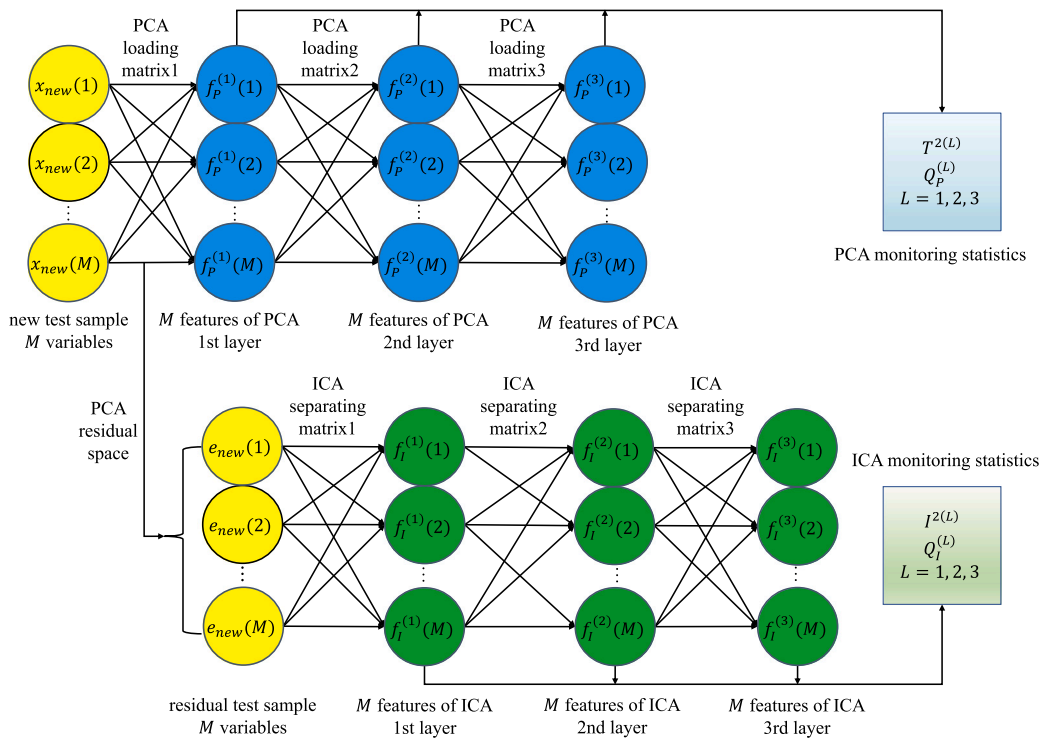

**Fig. 25.** **Computing monitoring statistics for a new test sample through DPI-3L.** The mapping matrices obtained in the offline training process are used to project the test sample into latent variables. Then monitoring statistics at each layer are built according to the extracted latent variables.

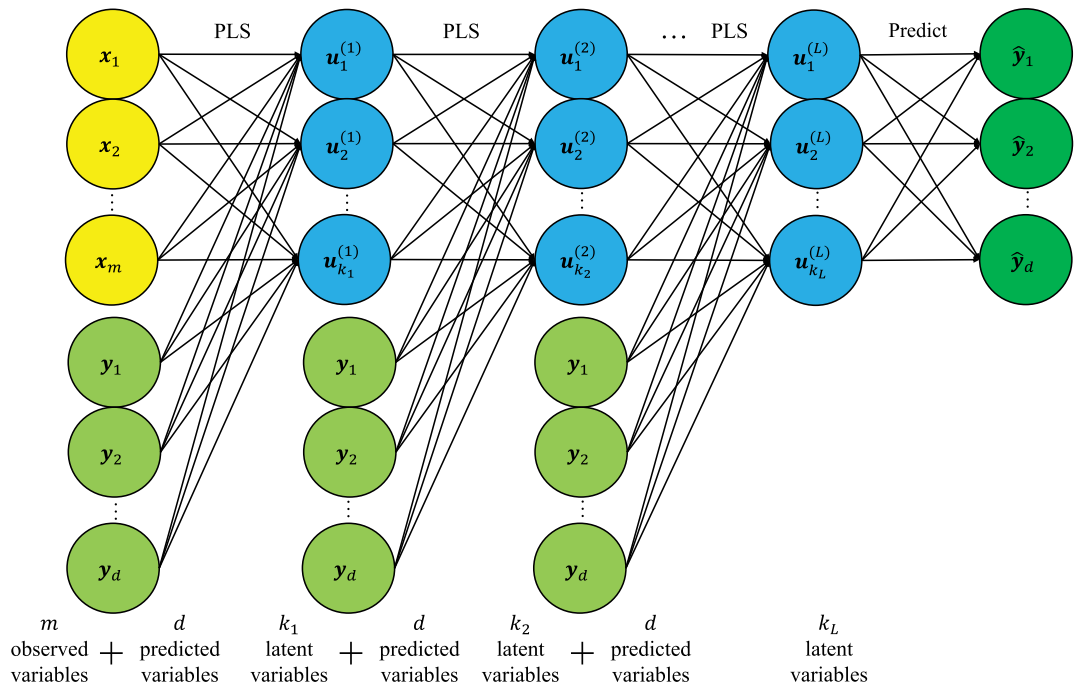

**Fig. 26.** **The model architecture of DPLS.** In the training phase, DPLS will perform matrix decomposition on the whole training set. In the testing phase, the test data will be passed forward through the entire model architecture.

such as low computation cost and satisfactory interpretability (Kong & Ge, 2022b).

### 5.3. Classification and regression based on LDLVM

This subsection generally refers to Kong and Ge (2022c). In this work, we provide a more general LDLVM that can be used for both classification and regression. Here we explore the concept of LDLVM from the perspective of partial least squares (PLS). We combine the virtues of deep model structure and PLS to propose a novel LDLVM called deep PLS, which is an interpretable and efficient supervised model. As a new kind of deep model, DPLS not only achieves competitive performance to DNN in appropriate application scenarios, but also has some unique advantages over DNN. The most obvious merits of DPLS over DNN are its *good interpretability*, *strong adaptability to data amounts*, *low computation complexity* and *short running time*.

### 5.3.1. Basic form of DPLS

The fundamental element in DPLS is the PLS model in Section 3.2.5. DPLS is a supervised model whose basic form is to use PLS to construct hierarchical and cascaded structures, then the obtained abstract representation in the highest layer is utilized to make predictions. The latent variables output by the current PLS layer are treated as inputs of next PLS layer, so that different layers are hierarchically connected by PLS. The architecture of an $L$-layers DPLS is illustrated in Fig. 26, where the superscript of the latent variable represents the layer number. When the predicted variables $[y_1, y_2, \ldots, y_d]$ are continuous values, DPLS is a regression model. If the predicted variables are discrete labels, DPLS can also be used to deal with classification tasks. In contrast to DNN, the connecting lines between each layer in Fig. 26 are slightly different. In a DNN, connecting lines typically represent the linear connection weights between two neurons, but in Fig. 26, they just stand for a type of association (e.g., the generation of $u_1$ depends on $[x_1, x_2, \ldots, x_m]$ and $[y_1, y_2, \ldots, y_d]$).

After giving the architecture of DPLS, we first prove its feasibility and merits theoretically. The goal of PLS, as mentioned in Section 3.2.5, is to maximize the covariance between $u_i$ and $v_i (1 \le i \le k)$. Here we shall theoretically prove that the deep structure of DPLS can increase the aforesaid covariance, and such a merit is described in Theorem 1.

**Theorem 1.** *The covariance between the first pair of latent variables gradually increases with the increase of layers in the DPLS model, i.e.*

$$\text{cov}(u_1^{(1)}, v_1^{(1)}) \le \text{cov}(u_1^{(2)}, v_1^{(2)}) \le \cdots \le \text{cov}(u_1^{(L)}, v_1^{(L)}). \tag{72}$$

The proof of Theorem 1 can be found in Kong and Ge (2022c).

The first latent variable has crucial significance. Since it is first extracted from the original data, it typically contains the greatest data information. Besides, the first latent variable determines the directions of other latent variables to a certain extent, as the remaining latent vectors need to be orthogonal to it. Furthermore, the covariance between the first pair of latent variables is typically an order of magnitude greater than that between the rest pairs of latent variables. As a result, the aforementioned *Theorem 1* — which demonstrates the relationship between the first pair of latent variables at different layers — is extremely important to the model structure of DPLS. In addition to the aforesaid merit, DPLS can also provide other virtues, such as *data visualization* and *variable importance assessment*, and more detailed discussions can be found in Kong and Ge (2022c).

### 5.3.2. Generalized form of DPLS

One may notice that though the aforesaid DPLS has clear theoretical merits, it is essentially a linear model. When the application scenarios are linear or approximately linear, DPLS may achieve good modeling results, but the performance of DPLS may not be satisfactory when the problems present strong nonlinearity. As a result, we provide a generalized form of DPLS in this subsection, which aims at enhancing the adaptability of DPLS so that it can handle nonlinear characteristics.

The first step is to modify the basic PLS model to a nonlinear form, and we propose an effective way to achieve this. Denote $\phi$ as a nonlinear mapping function that maps $X$ to a high-dimensional nonlinear space $H$ through $H = \phi(X)$. Then performing PLS decomposition on $H$ and $Y$ is able to introduce nonlinear factors into the modeling process. Detailed procedures of the nonlinear PLS (NPLS) algorithm can be found in Kong and Ge (2022c). After adding nonlinear mapping layers in the modeling process, we can get the generalized DPLS (GDPLS) model, which is shown in Fig. 27. The overall architecture of GDPLS is similar to that of DNN, which both can be understood as a deep model structure including linear connection layers and nonlinear mapping or activation layers. However, the forms of linear and nonlinear layers are quite different in GDPLS and DNN. The linear weights in DNN are acquired through random initialization and backpropagation-based optimization, whereas the linear layers of GDPLS are obtained through the PLS decomposition. Besides, GDPLS only has a forward process, while the training procedures of DNN involve iterative forward and backward process. Furthermore, another big difference lies in the nonlinear layers. The nonlinear activation function in neural networks must be differentiable, whereas a special advantage of GDPLS is it allows for either differentiable or non-differentiable mapping functions. The optimization of DPLS and GDPLS need not calculate the differentials since the proposed framework does not rely on backpropagation and gradient descent to train the model. As a result of this benefit, GDPLS can use various mapping functions, whether they are differentiable or not.

In addition to the flexibility of activation functions, another advantage of GDPLS over DNN is its lower computation complexity. Suppose $X \in \mathbb{R}^{n \times m}$, $Y \in \mathbb{R}^{n \times d}$ and $H \in \mathbb{R}^{n \times s}$. For a GDPLS, assume the number of extracted latent variables is $k$ and the number of layers is $L_1$, then the computation complexity of GDPLS is $\mathcal{O}(ns(s + kd)L_1)$. For a DNN with $L_2$ hidden layers and each layer contains $h_l (l = 1, \ldots, L_2)$ neurons, if the number of training epochs is $e$, then its complexity is $\mathcal{O}(enmd \prod_{l=1}^{L_2} h_l)$. In the training process of DNN, the term $e \prod_{l=1}^{L_2} h_l$ contributes a lot to the computations. In general, we have $\mathcal{O}(enmd \prod_{l=1}^{L_2} h_l) > \mathcal{O}(ns(s + kd)L_1)$ because of the term $e \prod_{l=1}^{L_2} h_l$.

Sections 5.3.1 and 5.3.2 not only provide the basic and generalized forms of DPLS, but also discuss their merits. In Kong and Ge (2022c), the superiority of deep PLS is evaluated on four practical examples, including two regression cases and two classification cases, where the proposed models achieve satisfactory performance, and simultaneously show good interpretability, high modeling efficiency and strong adaptability to data amounts.

## 6. Discussions and outlooks

We have provided a thorough overview on different types of LVMs and their industrial applications. Specifically, traditional LVMs have concise principles and good interpretability, but their model capacity cannot address complicated tasks. Benefiting from large amounts of training data and the breakthrough in computing power, neural networks-based DLVMs have sufficient model capacity to achieve satisfactory performance in complex scenarios, but it comes at a cost in terms of model interpretability and efficiency. Our proposed LDLVM combines the virtues of classic LVMs and DLVMs, which is able to achieve satisfactory results in suitable scenarios, and at the same time keeps a high level of interpretability and efficiency. There are still some additional details to be discussed around LVMs and their applications, also some promising outlooks deserving further investigations.

### 6.1. Other related models

Some other well-known LVMs are not included in this article, because they have not been extensively employed in industrial scenarios. Typical examples include latent Dirichlet allocation (LDA) (Blei, Ng, & Jordan, 2003), generative topographic mapping (GTM) (Bishop, Svensén, & Williams, 1998), etc. Moreover, there exist some models that used the idea of latent variables in their principles or modeling processes, but we do not completely regard them as typical LVMs. Popular examples are GAN, RNN and LSTM. GAN first assumes real data is generated by a latent variable $z$, usually a simple random noise, then trains a generator and a discriminator through an adversarial way, so that to gradually find the best mapping that projects the latent variable $z$ to practical data. Since the latent variable $z$ is usually chosen as Gaussian noise, and the most important part of a GAN is the generator, it is often considered as a generative model rather than an LVM. RNN and LSTM are used for modeling sequential data, and they both suppose the target in the current time is influenced by current features and the hidden state of previous time. The hidden states are essentially the latent variables in RNN and LSTM, but they are not assumed to be the underlying generators. Instead, they generally server as the middle factors for predicting the target. Therefore, RNN and LSTM are usually not considered as LVMs.

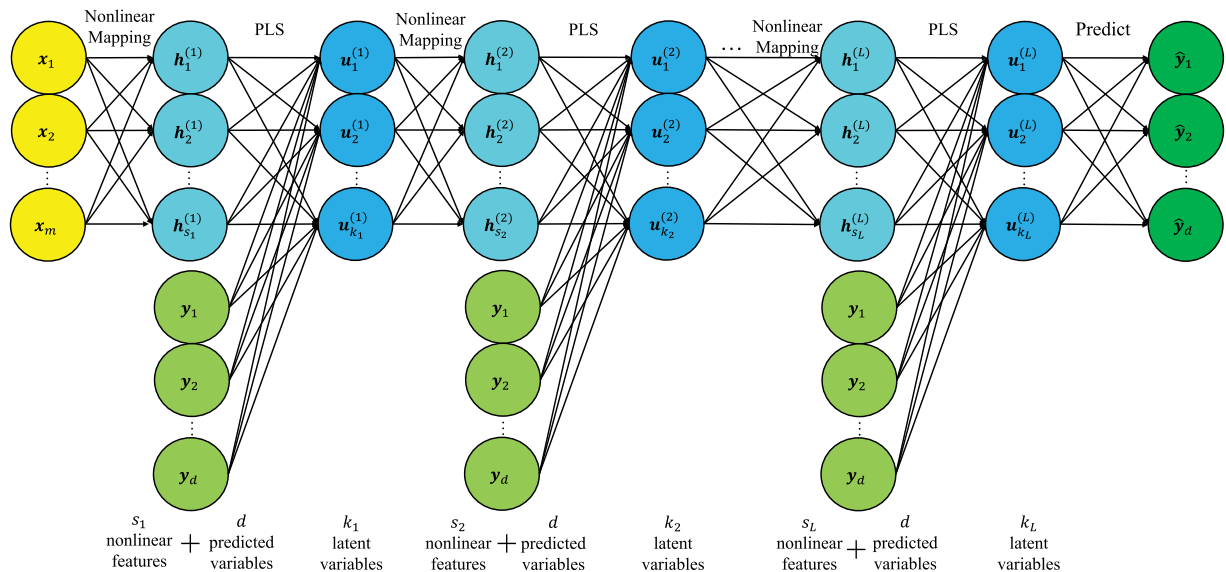

**Fig. 27.** The model architecture of GDPLS.

### 6.2. Extensions of the proposed two LDLVMs

The first LDLVM in Section 5.2 develops a novel framework for effectively deepening the traditional LVMs to improve their fault detection ability. Exploring suitable ways to utilize the proposed deep framework for other industrial scenarios, such as fault diagnosis, soft sensing, and so on, is worth investigating. The deep PLS in Section 5.3 is able to address linear and nonlinear classification and regression tasks, but the current form does not capture *unstructured* information well. How to apply deep PLS for unstructured data? How to effectively integrate spatial and temporal information in deep PLS? These problems need to be further studied to expand the application scope of the proposed framework.

### 6.3. Exploring the concept of LDLVM

LVMs are widely used in various areas, and the novel idea of LDLVM presented in Section 5.1 has powerful potentials in many application scenarios. We have successfully developed two LDLVMs that serve for different purposes. We believe the idea of LDLVM may open up a new avenue for DL, and deep models are no longer the antithesis of interpretability and efficiency. Under the concept of LDLVM, it is very promising to explore other ways to build transparent deep models with satisfactory performance, good interpretability and high efficiency. And this is especially meaningful for industrial scenarios, where we often need to balance model performance, efficiency, and reliability.

### 6.4. Incorporating prior knowledge into LVMs

LVMs are effective in capturing representative features from data, but the extraction process usually does not consider the relations, knowledge or other information between variables. Such materials are called *prior knowledge*, which exists in lots of real-world systems (Cheng, Wu, Tian, Wang, & Tao, 2020; Liu, Wu, Ge, Fan, & Zou, 2021; Niyogi, Girosi, & Poggio, 1998). In particular, prior knowledge exists, either explicitly or implicitly, in almost every industrial process. Such as the relationship between *valve opening* and *flow volume*, the relationship between *cooling water flow* and *equipment temperature*, etc. Integrating such information into the modeling process is able to improve the reliability and interpretability of LVMs, which may also enhance the performance of LVMs (Liu, Zeng, Xie, Luo, & Su, 2018; Zhai, Zeng, & Ge, 2021), since the extracted features have included practical constraints and are more in line with real scenarios.

### 6.5. Integration with data augmentation

Data is of great importance to intelligent algorithms, especially for deep models. In particular, for industrial systems, data issues have special significance. Since in many industrial scenarios, we may not be able to collect sufficient high-quality samples for data-driven modeling, which severely limits the performance of deep models. As a result, one purpose of LDLVM is to alleviate the dependence of deep models on data volume. Another effective way to mitigate this problem is *data augmentation*. Popular augmentation approaches include sample interpolation (Chawla, Bowyer, Hall, & Kegelmeyer, 2002), GAN (Frid-Adar et al., 2018; Jiang & Ge, 2021b), flow-based model (Dinh, Krueger, & Bengio, 2015), etc. The number of effective samples generated by data augmentation has an upper bound. When the amount of augmented samples is still unable to effectively train neural networks-based deep models, a combination of data augmentation and LDLVM may achieve more satisfactory performance. Besides, developing LDLVM-based lightweight data augmentation methods is also worth investigating. If the advantages of data augmentation and LDLVM can be adequately integrated, we may be able to overcome the data scarcity problem to a great extent.

### 6.6. Model security of LDLVM

Although the satisfactory results achieved by intelligent algorithms are inspiring, if the developed models cannot be *safely* applied, their actual deployment may be unsuccessful and unreliable (Taddeo, Mc-Cutcheon, & Floridi, 2019). Especially in the industrial scenarios. Industrial manufacturing usually has higher safety requirements than other areas. A little negligence could result in severe health and financial damages. Recently, adversarial machine learning has revealed that neural networks-based deep models are quite vulnerable to well-designed tiny perturbations (Miller, Xiang, & Kesidis, 2020), which means DNN has serious security problems when faced with adversarial attacks. As a new kind of deep model, LDLVM has obvious advantages over DNN in model efficiency and interpretability, but its security has not been investigated. Designing effective attack methods for LDLVM, evaluating its adversarial robustness, and comparing its security with DNN is another promising direction.

# 7. Conclusions

This article provides a thorough and comprehensive survey on latent variable models and their applications in industrial data-driven modeling. The contributions of this article can be summarized as follows:

1. We start by analyzing the characteristics of modern industry, whose features and developments make us seek suitable and powerful modeling methods. Then this article provides some preliminaries on ML and DL, in which we have also explained their relationship with LVM and DLVM.

2. We discuss the definition, theory and application of traditional LVMs in detail. The concept of LVM is first vividly illustrated, then we provide a comprehensive tutorial on classic LVMs and a brief survey on their industrial applications.

3. We present a thorough introduction to mainstream DLVMs with emphasis on the theory and model architecture. After that, we give a detailed survey of DLVMs' applications on two typical industrial areas, i.e. process monitoring and soft sensor.

4. An innovative concept, lightweight deep LVMs, is discussed in this article. We first elaborate the motivation and connotation of LDLVM, then provide two novel LDLVMs along with detailed descriptions of their principles, structures and advantages.

5. Important open questions and possible research directions are discussed in this article. We hope our insights would help guide the future research efforts in this field.

LVMs and their variants are widely employed in a variety of industrial scenarios, and they have excellent performance in data analytics, feature extraction and decision-making. We hope this article would be beneficial for summing up the history, examining the present, and looking forward to the future of this field, as well as potentially opening a new window for the researches on LVMs and even deep learning.

## Declaration of competing interest

The authors declare that they have no known competing financial interests or personal relationships that could have appeared to influence the work reported in this paper.

## Acknowledgments

This work was supported in part by the National Natural Science Foundation of China (NSFC) (92167106 and 61833014), and the Natural Science Foundation of Zhejiang Province (LR18F030001).